\documentclass[11pt, a4paper]{article}
\usepackage[T1]{fontenc}
\usepackage{a4wide, url}
\usepackage[english]{babel}
\usepackage{graphicx}
\usepackage{array}

\RequirePackage[authoryear]{natbib} 

\usepackage{amsmath}
\numberwithin{equation}{section}

\usepackage{mathtools}

\usepackage{setspace}
\usepackage{multirow}
\usepackage{amsthm}
\usepackage{amsfonts}
\usepackage{amssymb}
\usepackage{pgf} 
\usepackage{bm}
\usepackage{paralist}
\usepackage{pdflscape}
\usepackage{xfrac}
\usepackage[linesnumbered,ruled,vlined]{algorithm2e}
\SetKwInput{KwInput}{Input}                
\SetKwInput{KwOutput}{Output}              
\usepackage{enumitem,kantlipsum}
\usepackage{comment}
\usepackage{mathtools}
\usepackage{caption}

\usepackage{stackengine}
\stackMath
\newcommand\tenq[2][1]{%
\def\useanchorwidth{T}%
\ifnum#1>1%
\stackunder[0pt]{\tenq[\numexpr#1-1\relax]{#2}}{\scriptscriptstyle\thicksim}%
\else%
\stackunder[1pt]{#2}{\scriptscriptstyle\thicksim}%
\fi%
}

\usepackage{color}

\usepackage{comment}
\excludecomment{ext}\excludecomment{comment}

\usepackage[top=1in,bottom=1in,left=.75in,right=.75in]{geometry}

\usepackage{tabu}

\newcolumntype{L}[1]{>{\raggedright\let\newline\\\arraybackslash\hspace{0pt}}m{#1}}
\newcolumntype{C}[1]{>{\centering\let\newline\\\arraybackslash\hspace{0pt}}m{#1}}
\newcolumntype{R}[1]{>{\raggedleft\let\newline\\\arraybackslash\hspace{0pt}}m{#1}}

\DeclareMathAlphabet\mathbfcal{OMS}{cmsy}{b}{n}

\newcommand{\1}{\, {1\!\! {\mathbb I}}}

\newcommand{\mbf}{\mathbf}
\newcommand{\beq}{\begin{equation}}
\newcommand{\eeq}{\end{equation}}
\newcommand{\bea}{\begin{eqnarray}}
\newcommand{\eea}{\end{eqnarray}}
\newcommand{\ba}{\begin{array}}
\newcommand{\ea}{\end{array}}
\newcommand{\bit}{\begin{itemize}}
\newcommand{\eit}{\end{itemize}}
\newcommand{\ben}{\begin{enumerate}} 
\newcommand{\een}{\end{enumerate}}
\newcommand{\bpm}{\begin{pmatrix}}
\newcommand{\epm}{\end{pmatrix}}
\newcommand{\bbm}{\begin{bmatrix}}
\newcommand{\ebm}{\end{bmatrix}}

\newcommand{\pr}{^{\prime}}

\renewcommand{\l}{\left}
\renewcommand{\r}{\right}

\newcommand{\nn}{\nonumber}
\newcommand{\wh}{\widehat}

\newcommand{\n}{^{(n)}}

\newtheorem{ass}{Assumption}

\newtheorem{definition}{Definition}
\newtheorem{prop}{Proposition}

\graphicspath{{Figures/new/}}
\title{\textsc{\Large  \\ The  Dynamic, the Static, and the Weak \\
factor models and the analysis of high-dimensional time series
}}

\author{ Matteo {\sc Barigozzi} \\ Department of Economics\\ Universit\`a di Bologna, Italy\\  Email: matteo.barigozzi@unibo.it\\  $\,$ 
\\  Marc {\sc Hallin}\thanks{
    Marc Hallin gratefully acknowledges  the support of the Czech Science Foundation grants GA\v{C}R22036365 and GA24-10078S. } \\ ECARES and Department of Mathematics \\ Universit\' e libre de Bruxelles, Belgium \\   and \\    Institute of Information Theory and Automation\\  Czech Academy of Sciences, Prague, Czech Republic 
\\ Email: mhallin@ulb.ac.be\vspace{10mm}}

\date{ }

\begin{document}
\maketitle

\begin{abstract}  
Several  fundamental and closely interconnected  issues related to  factor models are reviewed and discussed: dynamic versus static loadings, rate-strong versus rate-weak factors, the concept of weakly common component recently introduced by \cite{GERSTH23}, 
  the irrelevance of cross-sectional ordering and the assumption of cross-sectional exchangeability,  the impact of undetected strong factors, and the problem of combining common and idiosyncratic forecasts. Conclusions all point to the advantages of the General Dynamic Factor Model approach of \citet{FHLR00} over the widely used Static Approximate Factor Model introduced by \citet{ChamberlainRothschild83}.
\end{abstract}

\bigskip

  \vspace{0.5cm}

  \vspace{0.5cm}

\noindent \textit{JEL Classification System:}  C32, C33, C53, C55.

\renewcommand{\thefootnote}{ } 

\thispagestyle{empty}

\renewcommand{\thefootnote}{\arabic{footnote}}


\section{Introduction: Spearman and the origins of factor models}  Factor models and factor model methods are rooted in   early-twentieth-century psychometrics.  It is usually admitted that the concept of {\it factor} first appears, more than a century ago, in  \cite{Spearman04} where a factor model is proposed in order to account for the dependencies between  several variables related with cognitive abilities   measured on  given individuals. The result was a two-factor theory in which  cognitive performance was explained by two unobservable  ``factors.''  The objective of these early factor models, thus, is to account for cross-sectional depen\-dencies---more specifically, cross-sectional covariances or correlations---in an i.i.d.\ context.

Spearman's exposition does not match the mathematical standards of present-day psychometrics or statistics, and more precise mathematical descriptions of factor models only came somewhat later. Classical references are, among many others, \cite{Hotelling33a,Hotelling33b}, \cite{Bartlett37, Bartlett38}, \cite{Joreskog69}, or the monograph by   \cite{LawleyMaxwell71};  see \cite{Joreskog07} for a historical account. 

In that i.i.d.~context, 
a factor model with $r$ factors is a statistical model characterized by an equation of the form (the index $t\in{\mathbb Z}$ here does not stand for time and is used to facilitate comparisons with the  time-series case to be described later)
\begin{equation}\label{1}
{\bf X}_t =\boldsymbol{\chi}_t + \boldsymbol{\xi}_t \coloneqq  {\bf B}{\bf f}_t + \boldsymbol{\epsilon}_t, \quad t= 1,\ldots T,
\end{equation}
%
%
where 
\begin{enumerate}
\item[(a)] ${\bf X}_1,\ldots,{\bf X}_T$ is an i.i.d.\   sample of $n$-dimensional observations ${\bf X}_t=(X_{1t},\ldots,X_{nt})^\prime$ with
(for ease of exposition and  without  loss of generality)
  ${\rm E}[X_{it}]=0$ and $0<{\rm E}[X_{it}^2]<\infty$, $i=1,\ldots, n$;
\item[(b)] ${\bf B }
$ is an unspecified  $n\times r$   matrix of scalar {\it loadings} with row vectors ${\bf B}_i=(B_{i1},\ldots,B_{ir})$, \linebreak $i=1,\ldots,n$;
\item[(c)] ${\bf f}_t=(f_{1t},\ldots ,{f}_{rt})^\prime$, $t=1,\ldots,T$ is an i.i.d.\  sample of  latent (unobservable) $r$-dimensional variables, the {\it (common) factors}, with ${\rm E}[{\bf f}_t]=\boldsymbol 0$ and ${\rm E}[{\bf f}_t{\bf f}_t\pr]= {\bf I}_{r\times r}$; call ${\boldsymbol \chi}_t=(\chi_{1t},\ldots,\chi_{nt})\pr\coloneqq{\bf B}{\bf f}_t$ the {\it common component}  of ${\bf X}_{t}$;
\item[(d)]  $\boldsymbol{\epsilon}_t=(\epsilon_{1t},\ldots ,\epsilon_{nt})^\prime$, $t=1,\ldots,T$  
is the unobserved realization of an  i.i.d.\  process of $n$-dimen\-sional   variables with  ${\rm E}[{\boldsymbol\epsilon}_t] = {\boldsymbol 0}$ and ${\rm E}[{\boldsymbol{\epsilon}}_t{\boldsymbol{\epsilon}}_t\pr]$ diagonal;  call ${\boldsymbol \xi}_t=(\xi_{1t},\ldots,\xi_{nt})\pr\coloneqq\boldsymbol{\epsilon}_t$ the {\it idiosyncratic component} of ${\bf X}_{t}$;
\item[(e)] the factors and the idiosyncratic component (hence also the common and the idiosyncratic components) are mutually orthogonal:
 ${\rm E}[{f}_{kt}\, {\epsilon}_{i t}]=0$,  $k = 1,\ldots, r$,  $i =1,\ldots, n$,  and $t =1,\ldots, T.$
\end{enumerate}
Equation \eqref{1} thus decomposes the component $X_{it}$ of the observation ${\bf X}_t$ into two unobservable and mutually orthogonal components: the  {\it common component} ${\chi}_{it} $ and the  {\it idiosyncratic component} ${\xi}_{it}$. 
 All subsequent factor models are based on such a decomposition, with various restrictions on $\chi_{it}$ and ${\xi}_{it}$: call it the {\it factor model decomposition}.
Because the idiosyncratic $ \boldsymbol{\xi}_t $ in \eqref{1}  is a vector of $n$ mutually orthogonal white noises, this traditional model is called  an {\it exact} factor model. The nature of \eqref{1}, along with assumptions (a)--(e), is that of a semiparametric statistical model with parameters the loadings~$\bf B$ and nuisances the unspecified joint distribution of ${\bf f}_t $ and $\boldsymbol{\epsilon}_t$. Note that, because of i.i.d-ness in (c) and (d), the orthogonality condition in (e) holds between ${f}_{kt}$ and  ${\epsilon}_{i t^\prime}$ for all $k = 1,\ldots, r$,  $i =1,\ldots, n$,  and $t =1,\ldots, T$, but also for all $t^\prime =1,\ldots, T$. If $t$ is interpreted as time, this means that (e) implies orthogonality between $\boldsymbol{\chi}_t$ and  $\boldsymbol{\xi}_t$ {\it at all leads and lags}; the same implication no longer  holds in subsequent sections, where $\boldsymbol{\chi}_t$ and  $\boldsymbol{\xi}_t$ are  time series exhibiting serial dependence. 

Estimation of this exact static factor model is typically performed via Gaussian MLE, which gives consistent (as $T\to\infty$) estimates  only for the loadings, while the factors (which can be treated as incidental parameters) cannot be consistently estimated since $n$ is   fixed \citep{AR56,LawleyMaxwell71,AFP87}.  The MLE, moreover, also requires finite fourth-order moments of the observables.

The observations in \eqref{1} being i.i.d., this exact model is of limited interest in the analysis of econometric data, where serial dependence is ubiquitous. The development of factor models in  time-series and econometrics  only started  in the late seventies with the {\it dynamic exact factor models} of \cite{Geweke77} and  \cite{SargentSims77}, who allow for factor-induced serial dependence  and, a few years later, with the {\it static approximate factor model} of   \cite{Chamberlain83}  and \cite{ChamberlainRothschild83}, who for the first time consider high-dimensional asymptotics (with the dimension $n$  tending to infinity) in the context. 

{The Chamberlain and Rothschild approach has been popularized  at the turn of the century thanks to the contributions of \cite{StockWatson02a, StockWatson02b}, \cite{BaiNg02}, and \cite{Bai03}, followed by many others. A few years earlier, \cite{FHLR00} and \cite{ForniLippi01} had introduced the {\it Generalized } or {\it General Dynamic Factor Model} (GDFM) which nests all other factor models. The static approximate  model of Chamberlain and Rothschild, however,   by far remains (with both $n$ and $T$ tending to infinity)  the most popular and widespread  form of factor model in applied econometrics.} 

{In Section \ref{StaticSec}, we briefly review the static approximate factor model of    \cite{Chamberlain83}  and \cite{ChamberlainRothschild83}. Section \ref{DynSec} deals with  the GDFM and, in particular, revisits \citet{HallinLippi13} by showing that the GDFM factor model decomposition is actually a representation result rather than just a statistical model. In Section \ref{Sec4} we reconsider the static approach of Chamberlain and Rothschild and show that, although less general than the GDFM, it can also be derived as a representation result; the proofs are developed in  the Appendix.  Section~\ref{weaksec}   discusses different concepts of weak factors, both static and dynamic. In Section~\ref{sec5}, we show, both empirically and theoretically, that  {\it weak factors}, in the usual acception of {\it rate-weak factors}, are incompatible with the a assumption of  cross-sectional exchangeability. 
Section~\ref{sec:idio}  examines the crucial role of idiosyncratic components in relation to undetected factors, and the importance, in forecasting problems,  of orthogonality  {\it at all leads and lags} between the common and  idiosyncratic components.  Section~\ref{concSec}  concludes. 
}

\section{The Static: 
Chamberlain 
 and 
 Rothschild
}\label{StaticSec}
%

The factor model decomposition proposed by \cite{
Chamberlain83} and \cite{ChamberlainRothschild83} with the objective of analizing high-dimensional financial time series reads
\begin{equation}\label{2}
{X}_{it} =
{\chi}_{it}^{\scriptscriptstyle\text{\rm{stat}}
} + 
{\xi}_{it}^{\scriptscriptstyle\text{\rm{stat}}
} \coloneqq    {\bf B}_i
{\bf f}_t   + 
{\xi}_{it}^{\scriptscriptstyle\text{\rm{stat}}
}
\quad i=1,\ldots , n,\ t= 1,\ldots T. 
\end{equation}
At first sight, this decomposition into  common and  idiosyncratic component   looks pretty much the same as the classical one \eqref{1}. A fundamental difference, however, is~that
 \begin{enumerate} 
 \item[(i)]the~$n\times 1$ observa\-tions~${\bf X}_1,\ldots,{\bf X}_T$ and the $r\times 1$ factors ${\bf f}_1,\ldots,{\bf f}_T$   no longer are assumed to be~i.i.d.,
 \item[(ii)]the idiosyncratic components ${\boldsymbol \xi}_t^{\scriptscriptstyle\text{\rm{stat}}} =({\xi}_{1t}^{\scriptscriptstyle\text{\rm{stat}}},\ldots, {\xi}_{nt}^{\scriptscriptstyle\text{\rm{stat}}})\pr$, $t=1,\ldots, T$   no longer are assumed to be~ i.i.d.\ white noise, and
 \item[(iii)]their covariance matrix needs not be diagonal. 
 \end{enumerate}
 Equation~\eqref{2}, thus, allows for serial cross-dependencies among the $X_{it}$'s, hence for the analysis\linebreak of  $n$-dimensional time series. 
   
Since  ${\boldsymbol \xi}_t^{\scriptscriptstyle\text{\rm{stat}}}$ is no longer white noise and can be (serially and cross-sectionally) correlated, the model is called an {\it approximate factor model} (as opposed to the {\it exact} ones). As a consequence, it  is not identified  for fixed $n$, even under the classical identification constraints. This motivates another crucial novelty:    high-dimensional asymptotics   are considered, with   both   $n$ and  $T$ going to infinity, where $n\to\infty$ yields {\it asymptotic} identification while $T\to\infty$ allows for consistent estimation as in the classical case. The loadings, however, remain  {\it static}  as in \eqref{1}---whence the terminology {\it static approximate factor model}.


Assumptions are quite mild: 
\begin{enumerate}
\item[  (a)] 
${X}_{11},\ldots {X}_{nT} $ is the observed  finite-$(n,T)$ realization of ${\bf X} =\{X_{it} \vert\, i\in{\mathbb N},\, t\in{\mathbb Z}\}$,  a second-order stationary process  with (for ease of exposition and  without  loss of generality) ${\rm E}[X_{it}]=0$\linebreak and~$0<{\rm E}[X_{it}^2]<\infty$, $i=1,\ldots,n$, $t=1,\ldots, T$;
\item[  (b)] ${\bf B }
$ is an unspecified  $n\times r$   matrix of scalar {\it loadings} with row vectors ${\bf B}_i=(B_{i1},\ldots,B_{ir})$, \linebreak $i=1,\ldots,n$;
\item[(c)] ${\bf f}_1,\ldots,{\bf f}_T$  is the unobserved  finite-$(n,T)$ realization of a second-order stationary~$r$-di\-mensional process ${\bf f}\coloneqq \{{\bf f}_t=(f_{1t},\ldots ,{f}_{rt})^\prime \vert\, t\in{\mathbb Z}\}$ of {\it factors}, with 
 ${\rm E}[{\bf f}_t]={\boldsymbol 0}$, and  ${\rm E}[{\bf f}_t{\bf f}_t\pr]={\bf I}_{r\times r}$;
\item[(d)] (d$_1$) letting $\chi_{it}^{\scriptscriptstyle\text{\rm{stat}} }\coloneqq {\bf B}_i{\bf f}_t$, the $n$-dimensional vector 
$\boldsymbol
{\chi}_{t}^{\scriptscriptstyle\text{\rm{stat}}
} \coloneqq (\chi_{1t}^{\scriptscriptstyle\text{\rm{stat}}
},\ldots ,\chi_{nt}^{\scriptscriptstyle\text{\rm{stat}}
})^\prime$ (the   {\it statically common} component) is the value at time~$t$ of the unobserved  finite-$(n,T)$ realization of a latent  second-order stationary process 
$$\boldsymbol{\chi}^{\scriptscriptstyle\text{\rm{stat}}}\coloneqq \{{\chi}_{it}^{\scriptscriptstyle\text{\rm{stat}}} \vert i\in{\mathbb N}, \  t\in{\mathbb Z}\};$$
(d$_2$)  denoting by ${\boldsymbol\Gamma}^{(n)}_{{\chi}}$ its  $n\times n$ lag-zero covariance matrix (which has rank $r$), the~$r$ nonzero eigenvalues of   ${\boldsymbol\Gamma}^{(n)}_{{\chi}}$ tend to infinity as $n\to\infty$;
\item[(e)] 
$\boldsymbol
{\xi}_{t}^{\scriptscriptstyle\text{\rm{stat}}
}=(\xi_{1t}^{\scriptscriptstyle\text{\rm{stat}}
},\ldots ,\xi_{nt}^{\scriptscriptstyle\text{\rm{stat}}
})^\prime$ (the   {\it statically idiosyncratic component})  is the value at time $t$ of the unobserved  finite-$(n,T)$ realization of a   second-order stationary process 
$$\boldsymbol{\xi}^{\scriptscriptstyle\text{\rm{stat}}}=\{{\xi}_{it}^{\scriptscriptstyle\text{\rm{stat}}} \vert i\in{\mathbb N}, \  t\in{\mathbb Z}\}$$
 with mean zero and  $n\times n$ lag-zero covariance matrix ${\boldsymbol\Gamma}^{(n)}_{{\xi}}$; as $n\to\infty$, all eigenvalues of   ${\boldsymbol\Gamma}^{(n)}_{{\xi}}$ remain bounded;
 \item [(f)] the factors ${\bf f}_t$ and the idiosyncratic components $\xi_{it}$ (hence also the common and the idiosyncratic components) are mutually orthogonal at time $t$:
 ${\rm E}[{f}_{kt}\, {\xi}_{i t}]=0$,  $k = 1,\ldots, r$,  $i\in\mathbb N$,  and $t \in
\mathbb Z.$
 \end{enumerate} 

  This static approximate factor model, like the  exact model \eqref{1}, constitutes a {\it semiparametric  statistical model}, with parameters  the loadings ${\bf B}_i$, $i =1,\ldots, n$ and nuisance the unspecified distributions of the idiosyncratic process $\boldsymbol\xi$ and the factors. Most people interpret the decomposition \eqref{2} (including the particular form of the common) as describing a data-generating process, which is irrelevant from a statistical perspective. Some  (e.g.  \citealp{Onatski12}, \citealp{BaiLi16}) even treat the values of the factors as parameters---i.e., condition their analysis on the unobserved latent factor process.
  
Chamberlain and Rothschild, however, do not include (d$_2$) in their assumptions (see their Definition~2); as a result,  the number of factors, in their version of \eqref{2}, is not well identified (that number  could be any $K\geq r$)\footnote{We thank Marco Lippi for attracting our attention to this fact.} 
and their factors need not be ``pervasive''  in the sense of inducing an unbounded (as~$n\to\infty$) ${\boldsymbol\Gamma}^{(n)}_{{\chi}}$ eigenvalue.   
 Nor do they impose any rate on the divergence, as $n\to\infty$,  of the diverging eigenvalues of the covariance matrix~${\boldsymbol\Gamma}^{(n)}_{{\chi}}$, which in principle can be sublinear, linear, or superlinear.  And they  do not consider the estimation problem for  their static approximate factor model. 
  
 Estimation
  is considered in  \cite{StockWatson02a}, \cite{Bai03}, and \citet{FGLR09}, among many others, who   provide a rigorous treatment of the asymptotic properties of  PCA-based estimators for the loadings and the factors of the Chamberlain and Rothschild  model and show that, as expected, consistency (up to orthogonal transformations, as usual) is achieved, as  both~$n$ and $T$ tend to infinity,  for the factors and the loadings.  
 {Finite fourth-order moments of the observables are also required. QMLE methods also have been considered under the same set of assumptions (\citealp{DGRqml,BaiLi16,BLqml}).}
  
Typically, once factors are {\it extracted} (or estimated, with a slight abuse of language when they are not considered as parameters) from an $n$-dimensional ($n$ large) time series ${\bf X}_t^{(n)}$ observed over $t=1,\ldots,T$, they are used in a second step to predict a given set of target variables \citep{StockWatson02a,StockWatson02b,baing06}. This approach, in general, offers sizeable improvements over univariate or low-dimensional (small-$n$) forecasting methods and, to this day, this  static approximate factor model --with (see Section~\ref{orthsec}) some variations on the orthogonality between common and idiosyncratic components---remains  {the most popular tool} in the analysis of high-dimensional time series.

\section{ The Dynamic: Forni, Hallin, Lippi, and Reichlin}\label{DynSec}

Before presenting the General Dynamic Factor   Model  proposed by \cite{FHLR00} and \cite{ForniLippi01}, let us consider the dynamic exact factor model of    \cite{Geweke77} and \cite{SargentSims77}, where the idea of dynamic loadings appears for the first time.

\subsection{A forerunner: the dynamic exact factor model of  Geweke, Sargent, and Sims
 }\label{31sec}

Five years before Chamberlain and Rothschild, \cite{Geweke77} and \cite{SargentSims77} had understood that, if factor models were to be used in  econometrics,  the time-series nature of econometric data could not be ignored. A major novelty in their approach  is based on the observation that factors, in economic time series, cannot be expected to be loaded simultaneously by {\it all} components of ${\bf X}$: typically, some components  will load ${\bf f}_t$ before some others. 
 Their model  accordingly considers {\it dynamic loadings}, where the row  vectors of scalar loadings ${\bf B}_i$ are replaced with row vectors ${\bf B}_i(L)$ of linear filters with square-summable coefficients, yielding    a  {\it dynamic exact}   factor model:  
  %
\cite{Geweke77}'s factor model decomposition takes the form 
\begin{equation}\label{3}
{X}_{it} =
{\chi}_{it}^{\scriptscriptstyle\text{\rm{dyn}}
} + 
{\xi}_{it}^{\scriptscriptstyle\text{\rm{dyn}}
} \coloneqq   {\bf B}_i(L)
{\bf f}_t   + 
{\xi}_{it}^{\scriptscriptstyle\text{\rm{dyn}}
}
\quad i=1,\ldots , n,\ t= 1,\ldots T.
\end{equation} 
 
 This idea of considering dynamic loadings was extremely innovative. On the other hand, Geweke, Sargent, and Sims do not go all the way with 
 taking into account the time-series nature of the data, as they  still assume an   {\it exact }  factor model, with i.i.d.\  and mutually orthogonal      idiosyncratic components. 
This is a terribly strong assumption, which cannot be expected  to hold in most econometric datasets. However, thanks to that assumption, their model is identified (up to an orthogonal transformation of the factors) and  traditional fixed-$n$ asymptotics can be considered.    The scope of their dynamic but exact factor model, thus,    is not  high-dimensional.    Finally, just as in Chamberlain and Rothschild, their approach  is a semiparametric   statistical modeling one. 

Variants of this dynamic exact approach have been proposed by several authors, among which \cite{EngleWatson81}, \cite{ShumwayStoffer82}, \cite{WatsonEngle83}, and \cite{QuahSargent93}, who  develop  a ``state-space approach''  where dynamics  are introduced by adding to the static factor model decomposition~\eqref{2} a parametric dynamic equation (typically, a VAR specification) for the factors. 
A high-dimensional generalization of the latter approach   was proposed by \citet{doz2011two,DGRqml}, which we do not discuss here.
%
%
%
\subsection{ The General Dynamic Factor Model}

The General Dynamic Factor Model (henceforth GDFM) proposed by  \cite{FHLR00} and \cite{ForniLippi01} is combining the dynamic loading features  of the exact factor model of \cite{Geweke77} and \cite{SargentSims77} with the flexibility of the  approximate  factor model of \cite{Chamberlain83} and \cite{ChamberlainRothschild83}. 
The following presentation  of the GDFM is inspired from the time-domain exposition of \cite{HallinLippi13} and slightly differs from the original ones in  \cite{FHLR00} and \cite{ForniLippi01}. By avoiding the spectral-domain approach of  \cite{FHLR00} and \cite{ForniLippi01}, it also avoids the assumption that a spectrum exists and the possibility of spectral eigenvalues diverging only on a subset of the frequency domain.

Differently from the previous ones in Sections~\ref{StaticSec} and~\ref{31sec}, the approach adopted here is {\it  purely nonparametric} (as opposed to semiparametric). The only fundamental assumption required for the existence of the GDFM factor model decomposition \eqref{6}   is that the observation, an~$n\times T$ panel,  is the finite   realization,  for~$1\leq i\leq n$ and $1\leq t\leq T$ ($n$ and $T$ ``large''), of 
 some double-indexed second-order time-stationary stochastic process 
\begin{equation}\label{5}
\mathbf{X}\coloneqq \{X_{it} \vert i\in\mathbb{N} ,  \  t\in\mathbb{Z}\}
\end{equation}
with (for convenience and without loss of generality) mean zero: ${\rm E}[X_{it}]=0$, $i\in \mathbb{N}$,  $t\in\mathbb{Z}$. 

Instead of being part of the model specification and imposing constraints on the data-generating process, the factor model decomposition \eqref{6} (see below) into common and idiosyncratic is canonical and ``endogenous,''  in the sense that any second-order time-stationary  process of the form \eqref{5} admits such a decomposition; and, when additional conditions are imposed, they only involve the distribution of the observations $X_{it}$ (typically, their    spectral density  eigenvalues). This is in sharp contrast with the previous  approaches by Chamberlain and Rothschild and Geweke, Sargent, and Sims,   where the  factor model decompositions \eqref{2} and \eqref{3} are part of the assumptions, along with conditions on their elements (typically, on the  eigenvalues of the common and idiosyncratic covariance matrices~${\boldsymbol\Gamma}\n_{\chi}$ and~${\boldsymbol\Gamma}\n_{\xi}$).  

Denote by $\mathcal{H}^{\bf X}$ the Hilbert space spanned by $\bf X$, equipped with the~L$_2$ covariance scalar product, that is, the set of all L$_2$-convergent linear combinations of~$X_{it}$'s ($(i,t)\in{\mathbb N}\times{\mathbb Z}$) and  limits of L$_2$-convergent sequences thereof. \smallskip 

{\begin{definition} {\rm A random variable $\zeta$  with values   
  in ${\mathcal H}^{\bf X}$ and variance $\sigma^2_\zeta$  is called    \textit{dynamically common}   if either   $\sigma^2_\zeta =0$ (hence $\zeta = 0$ a.s.)  or
   $\sigma^2_\zeta >0$   and~$\zeta/\sigma_\zeta$ is the limit  in quadratic mean, 
as~$n\to\infty$,  of a sequence of standardized elements of~${\mathcal H}^{\bf X}$  of the 
 form~$
 {w_{\bf X}
  \n}/\sigma\n_w$, where 
      $$w_{\bf X}
  \n\coloneqq \sum_{i=1}^n\sum_{t=-\infty}^\infty a_{it}^{(n)}X_{it},\quad  \text{with } \ \sum_{i=1}^n\sum_{t=-\infty}^\infty (a_{it}^{(n)})^2 =1\quad\text{for all $n\in{\mathbb N}$,}$$
and $\big(\sigma\n_w\big)^2\coloneqq {\text{Var}}(w
  _{\bf X}
  \n)$  is such that $\lim_{n\to\infty}\sigma\n_w
  =\infty$. 
  }
  \end{definition}\smallskip
 
 Note that the condition $\sum_{i=1}^n\sum_{t=-\infty}^\infty (a_{it}^{(n)})^2 =1$ for all $n\in{\mathbb N}$ in this definition can be replaced\linebreak  with~$0< C^-\leq \liminf_{n\to\infty}\sum_{i=1}^n\sum_{t=-\infty}^\infty (a_{it}^{(n)})^2 \leq \limsup_{n\to\infty}\sum_{i=1}^n\sum_{t=-\infty}^\infty (a_{it}^{(n)})^2\leq C^+ < \infty$, $n\in\mathbb N$, for some constants $C^-$ and $C^+$.

{\begin{definition} {\rm  Call    \textit{dynamically common space of $\bf X$}    
 the Hilbert space   $\mathcal{H}_{\scriptscriptstyle\text{\rm dyn com}}^{\bf X}$   spanned by the collection of all dynamically  common variables in ${\mathcal H}^{\bf X}$; call       \textit{dynamically  idiosyncratic  space}     its orthogonal complement (with respect to ${\mathcal H}^{\bf X}$)  $\mathcal{H}_{\scriptscriptstyle\text{\rm dyn idio}}^{\bf X}\coloneqq \big(\mathcal{H}_{\scriptscriptstyle\text{\rm dyn com}}^{\bf X}\big)^\bot$.}
  \end{definition}
  \noindent The spaces $\mathcal{H}^{\bf X}_{\scriptscriptstyle\text{\rm dyn com}}$ and   $\mathcal{H}^{\bf X}_{\scriptscriptstyle\text{\rm dyn idio}}$ always exist and are unique (one of them may reduce to $\{0\}$, in which case the other ones coincides with $\mathcal{H}^{\bf X}$); in particular, they do not depend on~$t$.  Intuitively, $\mathcal{H}_{\scriptscriptstyle\text{\rm dyn com}}^{\bf X}$ is the Hilbert space spanned by the exploding (as $n\to\infty$) normed linear combinations, for $i=1,\ldots,n$ and $t\in{\mathbb Z}$,  of the $X_{it}$'s,  and  the limits of convergent sequences thereof.   \medskip


  Projecting each $X_{it}$ onto  $\mathcal{H}^{\bf X}_{\scriptscriptstyle\text{\rm dyn com}}$ and its orthogonal complement $\mathcal{H}^{\bf X}_{\scriptscriptstyle\text{\rm dyn idio}}$  yields the canonical decomposition
 \begin{equation}\label{6}
 X_{it} = \chi_{it}^{\scriptscriptstyle\text{\rm dyn}} + \xi_{it}^{\scriptscriptstyle\text{\rm dyn}},\quad i\in\mathbb N,\ \ t\in\mathbb{Z}
 \end{equation}
of $X_{it}$ into a {\it dynamically common component} $\chi_{it}^{\scriptscriptstyle\text{\rm dyn}}$ and a {\it dynamically idiosyncratic} component $\xi_{it}^{\scriptscriptstyle\text{\rm dyn}}$, respectively, which by construction  are mutually orthogonal {\it at all leads and lags}---more precisely, $\chi_{it}^{\scriptscriptstyle\text{\rm dyn}}$ and  $\xi_{i^{\prime} t^{\prime}}^{\scriptscriptstyle\text{\rm dyn}}$ are mutually orthogonal for all $i,\,  t,\,  i^{\prime}$, and $t^{\prime}$. 

Call \eqref{6} the {\it General Dynamic Factor Model} (GDFM) {\it decomposition} of $\bf X$.    Contrary to \eqref{1} and~\eqref{2},  this decomposition    is  not imposed on the data; it  is canonical, always exists, and  (beyond second-order stationarity)  does not put any restriction on the data-generating process of ${\bf X}$. In that sense, it is not a statistical {\it model} involving {\it  parameters}, and constitutes a canonical representation result. Whether  that representation result is  describing a data-generating process or not is irrelevant from a statistical perspective.

This nature of \eqref{6} as a representation result was first emphasized in \cite{FHLR00} and  Forni and Lippi~(2001) where, however, an additional    frequency domain assumption (requiring the existence of a spectrum satisfying \eqref{7} below)  is imposed. So far, indeed, no assumption has been imposed on the second-order stationary process  $\bf X$. Adding the requirement that, for any $n\in\mathbb N$, ${\bf X}_t\n\coloneqq (X_{1t},\ldots, X_{nt})\pr$ admits an~$n\times n$
  {\it spectral density matrix} 
  $$\theta\mapsto {\boldsymbol\Sigma}\n_{ X}({\theta})\coloneqq \sum_{k=-\infty}^\infty {\rm E}[{\bf X}_t\n {\bf X}_{{t-k}}^{(n)\prime}]{\exp}(-{\iota} k\theta),\quad \theta\in (-\pi,\pi]$$  %
 ($\iota$ the imaginary root of -1) with eigenvalues  
$$\lambda\n_{X;1}(\theta)\ge \lambda\n_{X;2}(\theta)\ge \ldots   \ge  \lambda\n_{X;n}(\theta),\quad \theta\in (-\pi,\pi] $$ such that 
 \begin{equation}\label{7}
\lim_{n\to\infty}\lambda\n_{X;q}(\theta)=\infty\quad \text{ and }\quad \lim_{n\to\infty}\lambda\n_{X;q+1}(\theta)  <\infty, \quad \theta\text{-a.e. in }(-\pi, \pi]
\end{equation}
{for some finite  $q\in{\mathbb N}$ independent of $n$,} 
it can be shown \citep{ForniLippi01, HallinLippi13} that  the common component process  ${\boldsymbol\chi}^{\scriptscriptstyle\text{\rm dyn}}\coloneqq\{\chi_{it}^{\scriptscriptstyle\text{\rm dyn}} \vert\, i\in{\mathbb N},\, t\in\mathbb{Z}\}$  is  driven by a~$q$-dimensional ortho\-normal white noise process~$\{{\bf u}_t = (u_{1t} , \ldots , u_{qt})\pr \vert\, t\in\mathbb{Z}\}$ of  {\it common shocks}.  
The GDFM decomposition~\eqref{6}, in that case, takes the form 
 \begin{equation}\label{gdfmeq} X_{it} = \chi_{it}^{\scriptscriptstyle\text{\rm dyn}} + \xi_{it}^{\scriptscriptstyle\text{\rm dyn}}\eqqcolon    \sum_{i=1}^q {\bf B}_i(L)
{\bf u}_t   + \xi_{it}^{\scriptscriptstyle\text{\rm dyn}}
 \quad i\in\mathbb N,\ \ t\in\mathbb{Z}\vspace{-1mm}
 \end{equation}
 for  some  collection $ {\bf B}_i(L)\coloneqq  \left(
B_{i1}(L),\ldots , B_{iq}(L) \right)
$ of  
 linear 
   square-summable  filters  $B_{ij}(L)$, $j=1,\ldots , q$ for all 
 $i\in\mathbb{N}$, with $\{{\bf u}_t\vert\, t\in{\mathbb Z}\}$ the standardized innovations  of ${\boldsymbol \chi}^{\scriptscriptstyle\text{\rm dyn}} $.   These filters (for $i=1,\ldots,n$) 
  can be expressed in terms of the eigenvalues and eigenvectors of ${\boldsymbol\Sigma}\n_{ X}({\theta})$.  As a consequence, the GDFM  decomposition~\eqref{gdfmeq}    is 
  identified. 
 
Under  assumption \eqref{7}, the dynamically common space $\mathcal{H}_{\scriptscriptstyle\text{\rm dyn com}}^{\bf X}$ of $\bf X$, moreover, is spanned by the values (at    any time $t_0\in{\mathbb{Z}}$) of the $q$ first {\it dynamic principal components} $\{\psi_{jt}\vert\, t\in{\mathbb Z}\}$, $j=1,\ldots,q$  of~$\bf X$ (see \cite{Brillinger81} or \cite{Hallinetal18} for definitions). These dynamic principal components and their empirical counterparts play, in the GDFM and its estimation method as proposed in \cite{FHLR00}, a role similar to that of  traditional  principal components in the static factor model \eqref{2} of  \cite{Chamberlain83} and \cite{ChamberlainRothschild83} and its estimation as proposed in \citet{StockWatson02a}, \citet{Bai03}, etc. Being based on spectral density matrices, this technique  requires finite moments of order four, the existence of a spectrum, and the existence of an integer~$0\leq q<\infty$  such that \eqref{7} holds; note, however, that these assumptions, contrary to the assumptions (d) and (e) underlying \eqref{2}, only involve the observable process $\bf X$.  That technique, unfortunately, involves two-sided filters, hence performs poorly at the ends of the observation period---making it unsuitable in a forecasting context. An alternative estimation method involving only one-sided filters is proposed in \cite{FHLZ15} and studied in~\cite{FHLZ17} and~\cite{Barigozzietal24}.

It is important to stress that neither \cite{ForniLippi01} nor \cite{HallinLippi13} impose any rate on the divergence, as $n\to\infty$, of the eigenvalues of the spectral density matrix ${\boldsymbol\Sigma}\n_{ X}$. Indeed, no divergence rate assumption is needed for identifying the model as long as a diverging gap exists between the exploding and bounded eigenvalues. When turning to estimation, \cite{FHLR00} still do not impose any rates and, because of this,  cannot establish any consistency rates---on this latter issue, see \cite{FHLR04}. \cite{HallinLiska07}, \cite{FHLZ17}, and, in  a more general context nesting the GDFM,  \cite{BLVL23},  require the additional assumption of linear divergence rates, thus deriving also consistency rates for the estimators defined therein. This issue is discussed in Section \ref{weaksec}.

   \section{Orthogonality}\label{orthsec}
A central issue in the theory and practice of factor models is orthogonality between the  common and idiosyncratic components. The   GDFM literature  always naturally assumes that orthogonality between dynamically common and idiosyncratic components holds at all leads and lags, i.e., ${\rm E}[\chi_{it}^{\scriptscriptstyle\text{\rm dyn}}\, \xi_{jt'}^{\scriptscriptstyle\text{\rm dyn}}]=0$ for all~$i,\, j\in\mathbb N$  and $t,\,t' \in
\mathbb Z$ (see Section \ref{DynSec}).  In our canonical characterization~\eqref{6} of the GDFM  decomposition, this  orthogonality  at all leads and lags is not even an assumption and follows from the definition of $\mathcal{H}^{\bf X}_{\scriptscriptstyle\text{\rm dyn com}}$ and  $\mathcal{H}^{\bf X}_{\scriptscriptstyle\text{\rm dyn idio}}$.

On the other hand, a variety of orthogonality assumptions  can be found  in the literature on the static approximate factor model.\footnote{These assumptions, generally, are stated as orthogonality conditions between the factors $f_t$ and the idiosyncratic ${\xi}_{it}^{\scriptscriptstyle\text{\rm{stat}}}$ rather than between the  common component ${\chi}_{it}^{\scriptscriptstyle\text{\rm{stat}}}$ and the idiosyncratic ${\xi}_{it}^{\scriptscriptstyle\text{\rm{stat}}}$  but, since the loadings are deterministic, the two formulations are equivalent.}
\ben
\item [(A)]  ${\rm E}[\chi_{it}^{\scriptscriptstyle\text{\rm stat}}\, \xi_{jt}^{\scriptscriptstyle\text{\rm stat}}]=0$,  $i,\, j \in\mathbb N$  and $t \in
\mathbb Z.$, i.e., orthogonality between the common and idiosyncratic components 
 only holds contemporaneously. This is assumed in 
\cite{Chamberlain83}, 
\cite{ChamberlainRothschild83}, 
\citet{DeMoletal08,DeMoletal24},
\citet{fan2013large},
\citet{Gersingetal23}.  {And this is the way we characterize the static approximate factor model decomposition~\eqref{8}  in   Section \ref{Sec4}.} Then, as we show in Section \ref{Sec4}, and except for the assumption that  $\chi_{it}^{\scriptscriptstyle\text{\rm stat}}$ is the linear combination of a {\it finite} number $r$ of factors, the static approximate factor model decomposition of~$X_{it}$ into~$\chi_{it}^{\scriptscriptstyle\text{\rm stat}} + \xi_{it}^{\scriptscriptstyle\text{\rm stat}}$  is not a statistical model, but a representation result that always holds. 

\item [(B)] ${\rm E}[\chi_{it}^{\scriptscriptstyle\text{\rm stat}}\, \xi_{jt'}^{\scriptscriptstyle\text{\rm stat}}]=0$,  $i,\, j\in\mathbb N$  and $t,\,t' \in
\mathbb Z.$, i.e., orthogonality between common and idiosyncratic components  at all leads and lags. This is assumed, for instance,  in
\citet{FGLR09} and \citet{FL2024hallin}. Then the static approximate factor model is a {\it genuine statistical model}, imposing very strict  constraints on the data-generating process. This also makes the static a particular case of the dynamic one (up, perhaps, to some special cases where non-pervasive leads or lags of the factors are orthogonal to their present values). 

\item [(C)] $\{\chi_{it}^{\scriptscriptstyle\text{\rm stat}}\vert\,  i\in\mathbb{N},\, t\in{\mathbb Z}\}$ and $\{\xi_{it}^{\scriptscriptstyle\text{\rm stat}}\vert\,  i\in\mathbb{N},\,  t\in{\mathbb Z}\}$ are mutually independent processes. 
This, which is assumed in \citet{baing06}, \citet{doz2011two}, \citet{BLqml}, \citet{anderson2022linear}, and \citet{CCSVAR}, among others,  is a strong reinforcement of~(B).

\item [(D)] $\sup_{n, T\in{\mathbb N}} \max_{ i, j=1,\ldots, n} {\rm E}[ T^{-1}( \sum_{t=1}^T {\chi}_{it}^{\scriptscriptstyle\text{\rm{stat}}}{\xi}_{jt}^{\scriptscriptstyle\text{\rm{stat}}})^2]\le M$ for some finite $M>0$. 
 This is assumed by \citet{Bai03}, \citet{DGRqml}, and \citet{BaiLi16},\footnote{Note that \citet{StockWatson02a} do not formally require (D) but somehow  use it in their proofs.} who call it an assumption of   weak dependence between common and idiosyncratic components. 
 It requires finite fourth-order moments, some\-thing~(B) does not, and  
 is used for proving the consistency of  the sample covariance matrix as an estimator of its population counterpart, which in turn is needed to prove the consistency of PCA estimators of the $\chi_{it}^{\scriptscriptstyle\text{\rm stat}}$'s. 
  Intuitively close  to (B),  it is not directly comparable to (A), (B), or~(C). 

\item [(E)] Finally, some authors assume deterministic factors \citep{LawleyMaxwell71,baili12,Onatski12}. This approach is tantamount to conditioning on the factors or, equivalently, on the idiosyncratic components. Apparently, it solves the orthogonality issue since the $\chi_{it}$'s are no longer  random. But this is only an apparent solution: conditioning on $\boldsymbol\chi$, if orthogonality between~$\boldsymbol\chi$ and  $\boldsymbol\xi$ does not hold at all leads and lags, for instance, alters the mean of $\boldsymbol\xi$---the conditional expectation of $\xi_{it}^{\scriptscriptstyle\text{\rm stat}}$, typically, no longer is zero. Treating factors as deterministic quantities while maintaining the assumption that the $\xi_{it}^{\scriptscriptstyle\text{\rm stat}}$'s have expectation zero, thus, kind of brings back through the window an orthogonality assumption that was thrown out the door.
\een

As we shall see in Section~\ref{sec:idio}, these orthogonality assumptions (or their absence) play a major role in the problem of combining forecasts of the $\chi_{it}^{\scriptscriptstyle\text{\rm stat}}$'s and the  $\xi_{it}^{\scriptscriptstyle\text{\rm stat}}$'s into forecasts of the $X_{it}$'s.

\color{black}

%
%
%
%
%
%
%

\section{The  Static Approximate Factor Model revisited}\label{Sec4}  



Paralleling the definition  \eqref{6} of the GDFM decomposition in the previous section,  a representation-based characterization of the static approximate factor decomposition  of \citet{ChamberlainRothschild83} also can be obtained, which extends \eqref{2}. 

Throughout, let us assume that the process $\bf X$ is second-order stationary. Denoting by ${\mathcal H}^{{\bf X}_t}$ the Hilbert space spanned (at time~$t$) by ${\bf X}_t$, consider the following decomposition of ${\mathcal H}^{{\bf X}_t}$ into a   statically { (at time~$t$)} common   subspace ${\mathcal H}^{{\bf X}_t}_{\scriptscriptstyle\text{\rm stat com}}$ and its orthogonal complement (still at time $t$, and with respect to  ${\mathcal H}^{{\bf X}_t}$) 
 ${\mathcal H}^{{\bf X}_t}_{\scriptscriptstyle\text{\rm stat idio}}\coloneqq  ({\mathcal H}^{{\bf X}_t}_{\scriptscriptstyle\text{\rm stat com}})^{ \perp}$.\bigskip
 
 {\begin{definition} {\rm A random variable $\zeta$  with values  
  in ${\mathcal H}^{{\bf X}_t}$   and variance $\sigma^2_\zeta$  is called   \textit{statically  common-at-time-$t$}   if either   $\sigma^2_\zeta =0$ (hence $\zeta = 0$ a.s.)  or  $\sigma^2_\zeta >0$   and~$\zeta/\sigma_\zeta$ is the limit  in quadratic mean, 
as~$n\to\infty$,  of a sequence of standardized elements of~${\mathcal H}^{{\bf X}_t}$  
of the 
 form~$
 {w_{\bf X}
  \n}/\sigma\n_w$, where 
    \begin{equation}\label{w}
    w_{{\bf X}_t}
  \n\coloneqq \sum_{i=1}^n  b_{i}^{(n)}X_{it},\qquad  \text{with} \ \sum_{i=1}^n (b_{i}^{(n)})^2 =1 \quad\text{for all $n\in{\mathbb N}$,} \end{equation}
 is such that $\big(\sigma\n_w\big)^2\coloneqq {\text{Var}}(w
  _{\bf X}
  \n)$  is such that $\lim_{n\to\infty}\sigma\n_w
  =\infty$. }
 \end{definition}
 \smallskip
 
 Note that the condition $\sum_{i=1}^n (b_{i}^{(n)})^2 =1 $ for all $n\in{\mathbb N}$ in this definition can be replaced\linebreak  with~$0< C^-\leq \liminf_{n\to\infty}\sum_{i=1}^n (b_{i}^{(n)})^2 \leq \limsup_{n\to\infty}\sum_{i=1}^n (b_{i}^{(n)})^2\leq C^+ < \infty$, $n\in\mathbb N$, for some constants $C^-$ and $C^+$.

 {\begin{definition} {\rm  Call    \textit{statically  common  space  of $\bf X$ at time $t$}    
 the Hilbert space   $\mathcal{H}_{\scriptscriptstyle\text{\rm stat com}}^{{\bf X}_t}$   spanned by the collection of all statically  common (at time $t$) variables in ${\mathcal H}^{{\bf X}_t}$; call       \textit{statically idiosyncratic   space of ${\bf X}$ at time $t$}     its orthogonal complement (with respect to ${\mathcal H}^{{\bf X}_t}$)  $\mathcal{H}_{\scriptscriptstyle\text{\rm stat idio}}^{{\bf X}_t}\coloneqq \big(\mathcal{H}_{\scriptscriptstyle\text{\rm stat com}}^{{\bf X}_t}\big)^\bot$.}
  \end{definition}
 \noindent The spaces $\mathcal{H}_{\scriptscriptstyle\text{\rm stat com}}^{{\bf X}_t}$ and   $\mathcal{H}_{\scriptscriptstyle\text{\rm stat idio}}^{{\bf X}_t}$ always exist and are unique for all $t$.  Intuitively, $\mathcal{H}_{\scriptscriptstyle\text{\rm stat com}}^{{\bf X}_t}$ is the Hilbert space generated by the exploding (as $n\to\infty$) normed linear combinations of the $X_{it}$'s, $i=1,\ldots,n$, for given $t$. Contrary to its dynamic counterpart $\mathcal{H}_{\scriptscriptstyle\text{\rm dyn com}}^{\bf X}$,  it does  depend on~$t$. It readily follows from the definitions that, for all $t\in\mathbb Z$, $\mathcal{H}_{\scriptscriptstyle\text{\rm stat com}}^{{\bf X}_t}\subseteq \mathcal{H}_{\scriptscriptstyle\text{\rm dyn com}}^{\bf X}$.

 We then have the canonical  decomposition  
\begin{equation}\label{8}
 X_{it} = \chi_{it}^{\scriptscriptstyle\text{\rm stat}} + \xi_{it}^{\scriptscriptstyle\text{\rm stat}},\quad i\in\mathbb N,\ \ t\in\mathbb{Z}
\end{equation}
where $\chi_{it}^{\scriptscriptstyle\text{\rm stat}} $ and $\xi_{it}^{\scriptscriptstyle\text{\rm stat}}$ are the projections of $ X_{it}$ on $\mathcal{H}_{\scriptscriptstyle\text{\rm stat com}}^{{\bf X}_t}$ and $\mathcal{H}_{\scriptscriptstyle\text{\rm stat idio}}^{{\bf X}_t}$, respectively.  
In contrast with the dynamic decomposition~\eqref{6}, however,  orthogonality between the common-at-time-$t$ and the idiosyncratic-at-time-$t$  components $\chi_{it}^{\scriptscriptstyle\text{\rm stat}}$ and $\xi_{it}^{\scriptscriptstyle\text{\rm stat}}$   only holds at time~$t$---more precisely, $\chi_{it}^{\scriptscriptstyle\text{\rm stat}}$ and  $\xi_{i^{\prime} t}^{\scriptscriptstyle\text{\rm stat}}$ are mutually orthogonal for all $i,\,  i^{\prime}$, and $t$---while for $\chi_{it}^{\scriptscriptstyle\text{\rm dyn}}$ and $\xi_{it}^{\scriptscriptstyle\text{\rm dyn}}$ in \eqref{6} and  \eqref{gdfmeq}  orthogonality of $\chi_{it}^{\scriptscriptstyle\text{\rm dyn}}$ and $\xi_{i\pr t\pr}^{\scriptscriptstyle\text{\rm dyn}}$  holds  for all~$i,\,  i^{\prime}$,   $t$, and $t\pr$. 


Alone, \eqref{8} does not imply \eqref{2} along with conditions (a)--(f), though. Adding the requirement that the eigenvalues  
$$\lambda\n_{X;1} \ge \lambda\n_{X;2} \ge \ldots   \ge  \lambda\n_{X;n}  $$
of the $n\times n$ covariance matrices ${\boldsymbol\Gamma}^{(n)}_{{X}}$ of ${\bf X}\n_t\coloneqq (X_{1t},\ldots,X_{nt})\pr$ (which, in view of second-order stationa\-rity, do  not depend on $t$) are  such that
 \begin{equation}\label{9}
\lim_{n\to\infty}\lambda\n_{X;r} =\infty\quad \text{ and }\quad \lim_{n\to\infty}\lambda\n_{X;r+1}   <\infty
\end{equation}
for some finite  $r\in{\mathbb N}$ independent of $n$, 
it can be shown (see Proposition~\ref{Thmplus} below)  that $\mathcal{H}_{\scriptscriptstyle\text{\rm stat com}}^{{\bf X}_t}$ admits~$r$-dimensional orthonormal  bases ${\bf f}_t=(f_{1t},\ldots ,f_{rt})\pr$, any of which can be interpreted as the value at time~$t$ of an  $r$-tuple of factors, yielding  a Chamberlain and Rothschild static approximate  factor decomposition~\eqref{2} satis\-fying the assumptions  (a)--(f) of Section~2.1, with statically common component $\chi_{it}^{\scriptscriptstyle\text{\rm stat}}$ of the form~${\bf B}_i {\bf f}_t$.  




The process $\bf X$ being second-order stationary, it admits the canonical representation  \eqref{8}. Despite the fact that the Hilbert spaces~
$\mathcal{H}^{{\bf X}_t}$, $\mathcal{H}_{\scriptscriptstyle\text{\rm stat com}}^{{\bf X}_t}$, and   $\mathcal{H}_{\scriptscriptstyle\text{\rm stat idio}}^{{\bf X}_t}$ depend on~$t$, its~$n\times n$ covariance matrices~${\boldsymbol{\Gamma}}^{(n)}_{X}$,  their eigenvalues and the corresponding eigenvectors, do not;  
accordingly, the number $r$ of diverging eigenvalues in \eqref{9}, 
 their values and the corresponding eigenvectors, and the projection operators 
  from $\mathcal{H}^{{\bf X}_t}$ to $\mathcal{H}_{\scriptscriptstyle\text{\rm stat com}}^{{\bf X}_t}$ and~$\mathcal{H}_{\scriptscriptstyle\text{\rm stat idio}}^{{\bf X}_t}$  (which only depend on ${\boldsymbol{\Gamma}}^{(n)}_{X}$),  are the same for all $t$.

 The following result shows that \eqref{9} yields a characterization of the static approximate factor model which is much more parsimonious than the Chamberlain and Rothschild one; see  the Appendix   for a proof. 
 
 \begin{prop}\label{Thmplus}
 Let $\bf X$ be second-order stationary. Then, the following statements are equivalent:
 \begin{enumerate}
 \item[{\rm (A)}]  \eqref{9} holds, that is,   
$\lim_{n\to\infty}\lambda\n_{X;r} =\infty$ and $\lim_{n\to\infty}\lambda\n_{X;r+1}   <\infty$;
 \item[{\rm (B)}] \eqref{2}  holds and  satisfies {\rm (a)--(f)}, that is, $\bf X$ admits a static approximate factor model representation with $r$ factors  in the sense of Section~\ref{StaticSec}.
  \end{enumerate}
 Moreover, if (A)---hence also (B)---holds, then  \eqref{2}  coincides with \eqref{8}: namely,   $\chi_{it}^{\scriptscriptstyle\text{\rm stat}}$ and  $\xi_{it}^{\scriptscriptstyle\text{\rm stat}}$  in~\eqref{2}  and $\chi_{it}^{\scriptscriptstyle\text{\rm stat}}$ and  $\xi_{it}^{\scriptscriptstyle\text{\rm stat}}$ in  \eqref{8} coincide for all   $i\in\mathbb N$ and $t\in\mathbb{Z}$.

 \end{prop} 
 In other words, the 
  static approximate factor model with $r$ factors as defined in  Section~\ref{StaticSec} (a definition that, unlike the Chamberlain and Rothschild one,  includes (d$_2$))   is entirely characterized by  (A), i.e., by  the divergence   of $\lambda\n_{X;r}$ and the boundedness of $\lambda\n_{X;r+1}$ 
  as~$n\to\infty$---that is, by the existence of  a growing gap, in the spectrum of the covariance matrices~${\boldsymbol{\Gamma}}^{(n)}_{X}$ of   $\bf X$, between the $r$th and $(r+1)$th eigenvalues. 
 
 The advantage of   characterization (A) over
   characterization (B)  is not only its parsimony but, above all, the fact that it only involves features of the observable process~$\bf X$ while (B) is about the  unobservable common and idiosyncratic components of an  unobservable decomposition \eqref{2}. The desire for an observable-based characterization of their model is also very much present in    \cite{ChamberlainRothschild83}, all the more so that the unobservable decomposition  \eqref{2}, in their approach, is a postulated one.  
In their Theorem~4, they make a  step into the direction of such an observable-based characterization by showing that \eqref{9} implies the decomposition of ${\boldsymbol{\Gamma}}^{(n)}_{X}$ (for any~$n\geq r)$ into the sum of a covariance matrix with rank $r$   and a matrix with uniformly bounded (as $n\to\infty$) eigenvalues, hence that (A) implies   their definition (which requires (d$_1$) but not  (d$_2$): see their Definition 2) of a static approximate factor model. However,  because (d$_2$) is not included in that  definition, the converse does not hold, and 
that definition does not imply the divergence of $\lambda\n_{X;r}$, hence cannot imply (A). Therefore,  \cite{ChamberlainRothschild83} cannot state an equivalence result similar to Theorem~\ref{Thmplus}. 

This is, of course, a very minor logical issue, and adding    (d$_2$) to the Chamberlain and Rothschild definition  immediately restores its equivalence with (A). Nevertheless, and since their proof of Theorem~4  is scattered over several sections and involves some technical terminology specific of financial econometrics,
we are providing, in the Appendix, an independent and purely mathematical proof of Theorem~1.


Finally, it is important to note that neither \eqref{9} nor \citet{ChamberlainRothschild83}   impose any rate on the divergence, as $n\to\infty$,  of the eigenvalues of the covariance matrices~${\boldsymbol\Gamma}^{(n)}_{\chi}$ and ${\boldsymbol\Gamma}^{(n)}_{X}$, which in principle can be sublinear, linear, or superlinear.  Indeed, no divergence rate assumption is needed for identifying the model as long as an eigen-gap widening,  as $n\to\infty$, to infinity exists between bounded and unbounded eigenvalues. When turning to estimation, however, all works mentioned above (\citealp{StockWatson02a}, \citealp{Bai03}, \citealp{FGLR09}, etc.) make the additional assumption of a linear rate of divergence of the unbounded  eigenvalues, and  the same linear divergence assumption is made by \citet{BaiNg02} when proposing an information criterion to determine $r$  based on the behavior of sample eigenvalues (see Section~\ref{Sec7}). That linear divergence assumption is required for the consistency of the estimation methods, though, not by the identifiability of the number of factors.  We argue, in Section~\ref{sec5}, that such  linear divergence automatically follows under the quite natural additional  assumption of {\it cross-sectional exchangeability}.

\color{black}


%
%
%

\section{The  Weak: De Mol, Giannone, and Reichlin, Onatski, Hallin and Li\v{s}ka, and Gersing, Rust, and Deistler}\label{weaksec} 



The concept of {\it   weak factors} appears in various places in the literature with, however, quite diverse meanings. 
\begin{enumerate}
\item[(a1)] {  {\it Statically rate-weak factors}} in the static approximate factor model are introduced by \citet{DeMoletal08} and \citet{Onatski12}.
They are related with the eigenvalues $\lambda\n_{\chi ;j}$, $j=1,\ldots, r$, of the common component covariance matrix $\bm\Gamma_{\chi}^{(n)}$ which are  diverging at sublinear rates, usually parametrized as~$n^{\alpha _j}$ with~$\alpha _j <1$.  This is the most popular setting, see   \citet{lam2012factor}, \citet{Freyaldenhoven22},  \citet{uematsu2022estimation}, and \citet{BaiNg23}, among many others.

\item[(a2)] {  {\it Dynamically rate-weak factors}} in the GDFM are considered by
\citet{BF24} who propose an estimator of a low-rank plus sparse spectral density matrix, under the assumption that the low-rank component  has eigenvalues $\lambda\n_{\chi;j}(\theta)$, $j=1,\ldots,q$, diverging at sublinear rates~$n^{\alpha_j}$ with~$\alpha_j <1$, $\theta\text{-a.e. in }(-\pi, \pi]$.

\item[(b)] {\it Block-specific factors} are introduced by \cite{HallinLiska11}, under a GDFM approach, in the context of panels divided into subpanels or blocks. The terminology {\it weakly common} factors is used for dynamic factors that are dynamically common in some block(s) but dynamically  idio\-syncratic in some other(s). The associated spectral eigenvalues are assumed to be diverging with the dimension of the block but no rate is specified. Static approximate  factor models with a hierarchical block structure are considered, for example,  by \citet{MNP13} but without any explicit mention of weak factors.

\item[(c)] Finally, \cite{GERSTH23} and \cite{Gersingetal23} define {\it   weakly common components} as {  the dif\-ference} (at time $t$) {  between the dynamically common  and the statically common components} of~$X_{it}$. Any (possibly infinite-dimensional) orthonormal basis of the space they are spanning (at time $t$) then can be considered a basis of {\it weak factors at time $t$}. This notion of weakness bears no relation to (a) and (b) above:   to avoid confusion, call   {\it GRD-weak} these weak factors. 
\end{enumerate}

\subsection{Rate-weak factors: De Mol, Giannone and Reichlin, Onatski}\label{SecRateW}

Following \citet{DeMoletal08} and \citet{Onatski12}, we will call a static or dynamic factor {``rate-weak''} or  {``weakly influential''} if the corresponding loadings, as~$n\to\infty$, are too small, or too sparse, for the corresponding eigenvalues to diverge at rate~$n$. 

As mentioned in the previous sections, as long as we have a widening, as $n\to\infty$, of the eigen-gap between the first $q$ or $r$ eigenvalues of the spectral density or covariance matrix of the observables~$\mbf X\n$ and the  remaining $n-q$ or $n-r$ ones, the GDFM or the  static approximate factor models are identified. Thus, for identification purposes, there is no need to assume any specific divergence rate for the divergent  eigenvalues.

However, in estimation, either via dynamic or static PCA, it is often assumed that eigenvalues diverge linearly. Under this assumption, the consistency rates for the  estimated common components derived in the literature  always involve two terms: the first one is the classical $\sqrt T$ term related to estimation of the covariance or the spectrum (in this latter case, the rate also depends  on the chosen bandwidth), the second one is related to the identification of the model, and thus to the rate of divergence of the eigengap, which under linear divergence yields a rate which is $\sqrt n$. Inspection of the proofs easily shows  that if the divergence rate were superlinear, consistency simply would  be achieved at a faster rate. 

Conversely, if the divergence were sublinear---rate $n^\alpha$, say,  with $\alpha<1$---consistent estimation becomes non-trivial. Indeed, as shown by \citet{Freyaldenhoven22} in the static  case, PCA only allows us to recover the factors which are associated with eigenvalues with divergence   rate a least equal to $\sqrt n$, i.e., rate $n^\alpha$ with $\alpha>1/2$ (see also \citealp{Onatski12} and recent work by \citealp{fan2024can}). A similar result can be easily derived for dynamic PCA.

As we shall argue in Section~\ref{sec5}, however, nonlinear growth rates, whether for static or for dynamic eigenvalues, are incompatible with the principle of  irrelevant cross-sectional ordering. The question is then: for which kind of data is cross-sectional ordering relevant and thus compatible with rate-weak factors? We refer to \citet{Freyaldenhoven22} for a list of examples, while here we just notice that they almost all fall into the category of data having an intrinsic block structure. Such block structure can be either  (1) known a priori, or (2) unknown. Case (1) is discussed in the next section. We do not discuss case (2) here, since  it resembles case (1) for most aspects but   estimation:   the block structure  being latent,  indeed, classical PCA no longer applies and alternative approaches are required, see, e.g., \citet{ando2017clustering}. 

%
%

\subsection{Weak factors in panels with block structure: Hallin and Li\v ska}\label{HalLisSec}

Consider a panel with a $K$ blocks---namely, an $(n\times T)$ panel divided into $K$ $(n_k \times T)$ subpanels with~$n=\sum_{k=1}^K n_k$; for asymptotics, assume that all $n_k$'s tend to infinity as $n,T\to\infty$. 

In a GDFM context, \citet{HallinLiska11} show that   the Hilbert space~${\mathcal H}^{\bf X}_{\scriptscriptstyle\text{\rm dyn com}}$ then decomposes into a  direct sum involving a {\it strongly common} 
  subspace ${\mathcal H}_{\scriptscriptstyle\text{\rm dyn com strong}}^{\bf X}$ spanned by factors which are common for each $n_k$-dimensional subpanel, plus $\sum_{j=1}^{n-1} {n\choose j}$ mutually orthogonal 
 {\it weakly common} sub\-spaces~${\mathcal H}_{\scriptscriptstyle\text{\rm dyn com weak}}^{\mbf X, j}$
    spanned by factors which are common for a  $j$-tuple of subpanels but   idiosyncratic for the other ones. {\it Strongly} and {\it weakly common}, here, is not related to any divergence rate, and should be understood as ``strongly/weakly common in the Hallin-Li\v{s}ka sense". 
    That decomposition is a refinement of the {\it global} dynamic factor model decomposition~\eqref{gdfmeq}.
  
 Hallin and Li\v{s}ka do not discuss rate-strength issues, but such issues naturally enter into the picture when the  relative magnitudes $n_k/n$ of the $K$ block sizes exhibit different asymptotic behaviors.  It is easy to see that  that   a given block $k_0$ may contain factors that are rate-strong within the block (yielding~$n_{k_0}$-linear divergence rate),  but   globally rate-weak (with  $n$-sublinear divergence rate). This happens if, for instance, that factor is idiosyncratic in all other blocks and~$n_{k_0}/n\to 0$ as $n_{k_0},n\to\infty$.    There is little reason, however, to assume that  $n_{k_0}/n\to 0$ for some block $k_0$, as this choice of an asymptotic scenario belongs to the analyst (the observation provides no information on the asymptotic behavior of $n_{k_0}/n$). The objective of that choice is a good asymptotic approximation of the actual, finite-$n$ situation, and the actual value of $n_{k_0}/n$ is likely to provide a better approximation than an asymptotic value arbitrarily set to zero. 
  
  \color{black}
  
 A similar approach could be taken, {\it mutatis mutandis},  with similar comments, in the context of  static approximate factor models, yielding a refinement of the static factor model decomposition \eqref{2}. 


\subsection{{\it\bf Weak factors are everywhere: Gersing, Rust, and Deistler}}

Under the intriguing title {\it ``Weak factors are everywhere,\!''} \cite{Gersingetal23}, with the objective of   {\it  reconciling} the static and the dynamic  approaches,  show (their  definitions slightly differ from the ones considered here)  that the statically common  space   $\mathcal{H}_{\scriptscriptstyle\text{\rm stat com}}^{{\bf X}_t}$, for any~$t$, is a subspace of the dynamically common one $\mathcal{H}_{\scriptscriptstyle\text{\rm dyn com}}^{{\bf X}}$ and call {\it weakly common} at time $t$ the elements of the difference~$\mathcal{H}_{\scriptscriptstyle\text{\rm dyn com}}^{{\bf X}}\setminus \mathcal{H}_{\scriptscriptstyle\text{\rm stat com}}^{{\bf X}_t}$: we shall call  them {\it GRD-weak}.

Let $\bf X$ be second-order stationary and satisfy both \eqref{7} and \eqref{9}. This result by \cite{Gersingetal23}  naturally yields a   new canonical factor decomposition, now involving three terms, of the form \vspace{-3mm}
 \begin{align}\nonumber
 \phantom{ X_{it}} &\phantom{= {\chi_{it}^{\scriptscriptstyle\text{\rm stat}}  \phantom{+\chi_{it}^{\scriptscriptstyle\text{\rm weak}}}} \hspace{2mm}} \raisebox{-3mm}{$\hspace{4mm}\xi_{it}^{\scriptscriptstyle\text{\rm stat}}$ }\\ 
  X_{it} &= \underbrace{\chi_{it}^{\scriptscriptstyle\text{\rm stat}} \hspace{-4mm} \phantom{\hspace{4mm}+\chi_{it}^{\scriptscriptstyle\text{\rm weak}}}} \hspace{-10mm}+ \ \overbrace{{  \chi_{it}^{\scriptscriptstyle\text{\rm weak}}} + \  \xi_{it}^{\scriptscriptstyle\text{\rm dyn}}}, \qquad i\in{\mathbb N},\ t\in{\mathbb Z} \label{10}\\
   \phantom{ X_{it}} &\phantom{= {\chi_{it}^{\scriptscriptstyle\text{\rm stat}} }}
  \raisebox{2mm}{ ${\chi_{it}^{\scriptscriptstyle\text{\rm dyn}} }$} \vspace{-3mm}\nonumber
\end{align} 
where $  \chi_{it}^{\scriptscriptstyle\text{\rm weak}} $---the {\it GRD-weakly common component} of~$X_{it}$---is the difference between the dynamically and statically common components $\chi_{it}^{\scriptscriptstyle\text{\rm dyn}} $ and $\chi_{it}^{\scriptscriptstyle\text{\rm stat}} $ (equivalently, the difference between the statically and dynamically idiosyncratic components $\xi_{it}^{\scriptscriptstyle\text{\rm stat}} $ and $\xi_{it}^{\scriptscriptstyle\text{\rm dyn}} $)  of $X_{it}$. 
Denoting by ${\mathcal H}^{{\bf X}_t}_{\scriptscriptstyle\text{\rm weak}}$ the Hilbert space generated by $\mathcal{H}_{\scriptscriptstyle\text{\rm dyn com}}^{{\bf X}}\setminus \mathcal{H}_{\scriptscriptstyle\text{\rm stat com}}^{{\bf X}_t}$ (which depends on $t$) , we have, for all $t\in{\mathbb Z}$, the inclusions~${\mathcal H}^{{\bf X}_t}_{\scriptscriptstyle\text{\rm weak}}\subseteq ~\!\mathcal{H}_{\scriptscriptstyle\text{\rm stat idio}}^{{\bf X}_t}$ and~${\mathcal H}^{{\bf X}_t}_{\scriptscriptstyle\text{\rm weak}}\subseteq \mathcal{H}_{\scriptscriptstyle\text{\rm dyn com}}^{{\bf X}}$, hence
\begin{equation} \label{GersingHilbert}
\mathcal{H}_{\scriptscriptstyle\text{\rm stat com}}^{{\bf X}_t}
{\perp}_{t}\,\, 
{\mathcal H}^{{\bf X}_t}_{\scriptscriptstyle\text{\rm weak}}
\perp
 \mathcal{H}_{\scriptscriptstyle\text{\rm dyn idio}}^{{\bf X}},\quad t\in{\mathbb Z}
\end{equation}
where $\perp$  and ${\perp}_t$ mean orthogonal at all leads and lags and orthogonal at time $t$, respectively.

This is a very ingenious finding, which shows how the GDFM (as defined in \eqref{gdfmeq}) is overarching the static model  (as defined in \eqref{2}) and, with the additional assumption \eqref{9}, strictly nests   the popular  static approximate model of  \cite{Chamberlain83} and \cite{ChamberlainRothschild83}. The fact that  the dynamically common space $\mathcal{H}_{\scriptscriptstyle\text{\rm dyn com}}^{{\bf X}}$ incorporates the GRD-weakly common components which, in the static approach, are idiosyncratic     explains the empirical finding \citep{FGLS18} that the GDFM performs better in terms, e.g.,  of forecasting, than the static model~\eqref{2} of Chamberlain and Rothschild even when the assumptions of the latter are satisfied. See Section~\ref{sec:idio} for major further   consequences of~\eqref{10} and the orthogonality properties~\eqref{GersingHilbert} on the relative  forecasting performance of the static and the GDFM approaches.

Note that, from our definitions of the dynamically common  and statically common-at-time-$t$\linebreak  spaces~$\mathcal{H}_{\scriptscriptstyle\text{\rm dyn com}}^{{\bf X}}$  and  
   $\mathcal{H}_{\scriptscriptstyle\text{\rm stat com}}^{{\bf X}_t}$ (which slightly differ from the \cite{Gersingetal23} ones), the inclusion~$\mathcal{H}_{\scriptscriptstyle\text{\rm stat com}}^{{\bf X}_t}\subseteq \mathcal{H}_{\scriptscriptstyle\text{\rm dyn com}}^{{\bf X}}$  trivially holds for any $t\in\mathbb{Z}$  so that, under  the unique  assumption of second-order stationarity,  $X_{it}$ for all $i$ and all~$t$ admits   the canonical decom\-position~\eqref{10}.

  In this  canonical decomposition \eqref{10}, $\chi_{it}^{\scriptscriptstyle\text{\rm weak}}$, being in the dynamically common space $\mathcal{H}_{\scriptscriptstyle\text{\rm dyn com}}^{{\bf X}}$,   is orthogonal, at all leads and lags, to the dynamically idiosyncratic component ${\xi_{it}^{\scriptscriptstyle\text{\rm dyn}} }$; being in the statically idiosyncratic-at-time-$t$ space, $\chi_{it}^{\scriptscriptstyle\text{\rm weak}}$ is also orthogonal to the statically common-at-time-$t$ component  ${\chi_{it}^{\scriptscriptstyle\text{\rm stat}} }$ (but not necessarily  its leads and lags); that ${\chi_{it}^{\scriptscriptstyle\text{\rm stat}} }$ might even be a.s.\ zero (yielding a statically purely idiosyncratic process). Therefore, $\chi_{it}^{\scriptscriptstyle\text{\rm weak}}$ cannot be a static factor, neither rate-strong nor rate-weak, and  typically consists of non-pervasive leading or   lagging values  of the (rate-weak or rate-strong) dynamic factors 
  or combinations thereof.  This  concept of GRD-weakly common component, thus, is unrelated to the  concepts of  rate-weak factors discussed in Sections~\ref{SecRateW} and \ref{HalLisSec}.

\section{ Exchangeability: weak factors are nowhere!}\label{sec5}

When no block structure is present in the data, one may wonder whether the concept of rate-weak factors is really relevant in practice. 

An all too often overlooked feature of panel data  is that, unlike  time-ordering,  their cross-sectional ordering is entirely arbitrary. Very often, some alphabetical ordering is adopted,  a choice that  should not have any impact on the analysis. An observed  panel, actually, should be treated as  the equivalence class of  its $n!$ cross-sectional permutations and all statistical procedures, all forecasts, all inferential conclusions,  should be invariant or equivariant under these cross-sectional permutations. 

In this section,  for simplicity of notation, we only consider the static case. The same arguments can be made verbatim for the dynamic case at the price of heavier notation.

\subsection{Factors and cross-sectional permutations: intuition}

To start with, let us conduct a few numerical exercises to show how cross-sectional ordering is related to the growth of eigenvalues, hence their divergence rates. First, we consider a panel generated from the very simple  static one-factor model 
\begin{equation}\label{11}X_{it} = B_if_t + \varepsilon_{it}\qquad i=1,\ldots,n_0,\  t=1,\ldots,T,
\end{equation}
where $\{f_t\vert\, t\in{\mathbb Z}\}$  and  $\{\varepsilon_{it}\vert\, t\in{\mathbb Z}\}$, $i=1,\ldots,n_0$, 
are $(n_0+1)$ mutually orthogonal unit variance white~noise processes, and the $n_0$-dimensional  column vector of loadings $\mbf B= (B_1,\ldots, B_{n_0})^\prime$ is of the form $\mbf B=  (\mbf A_1^\prime, \mbf A_2^\prime, \mbf A_3^\prime)^\prime$, with ${\mbf A}_1=0.25\1$, ${\mbf A}_2=\1$, and ${\mbf A}_3= c\1$, with $\1$ a $n_0/3$ column vector of ones and $c$ scaled such that $\mbf B^\prime \mbf B/n_0=1$. Here $n_0=240$ and $T=100$, hence $\mbf A_1$, $\mbf A_2$, and $\mbf A_3$ are~$80$-dimensional.

 The left-hand plot in Figure~\ref{Fig1} shows the evolution, for increasing $n=1,\ldots, n_0$, of the first eigen\-value~$\lambda_{X;1}^{(n)}$ (normalized by $\lambda_{X;1}^{(n_0)}$) of the~$(n\times n)$ lag-zero covariance matrix ${\boldsymbol\Gamma}^{(n)}_{X}$ in  the case of three cross-sectional permutations of the $X_{it}$'s.
Let $B_{(i)}$, $i=1,\ldots, n_0$ denote the $i$th order statistic of the loadings: $B_{(1)}\leq B_{(2)}\leq~\!\ldots~\!\leq~\!B_{(n_0)}$. The first permutation (dashed line) is ordering the cross-sectional items by increasing values of the loadings ($B_i=B_{(i)}$); the second permutation  (dashed-dotted line)  orders them  by decreasing values ($B_i=B_{(n_0-i+1)}$); the third  one (solid line)    is alternating small and large values of  the same ($B_1=B_{(n_0)}, B_2=B_{(1)}, B_3=B_{(n_0-1)},\ldots,B_{n_0}=B_{(n_0/2)}$ for even $n_0$). From a na\"\i ve point of view, one would be tempted by a superlinear extrapolation  in the first case (hence, a rate-superstrong factor) and  a sublinear one (a rate-weak factor) in  the second case, whereas a linear extrapolation (a rate-strong factor) looks quite reasonable in the third case. The three cases, however, are dealing with the same panel.
  
Second, we consider a panel generated from the static one-factor model
\begin{equation}\label{11rand}
X_{it,m} = (B_{i} u_{i}) f_t + \varepsilon_{it}\qquad i=1,\ldots,n_0,\  t=1,\ldots,T,
\end{equation}
where  $\{f_t\vert\, t\in{\mathbb Z}\}$,  $\{\varepsilon_{it}\vert\, t\in{\mathbb Z}\}$, and $B_i$ 
 are  generated as in \eqref{11}, and $u_{i}\sim N(0,1)$, $i=1,\ldots, n_0$. 
In the right-panel of Figure~\ref{Fig1} we show  the evolution for increasing $n=1,\ldots, n_0$ of the first eigenvalue $\lambda_{X;1}^{(n)}$ (normalized by $\lambda_{X;1}^{(n_0)}$) of ${\boldsymbol\Gamma}^{(n)}_{X}$ under the following permutations. 
First  (dashed line), when ordering the cross-section by increasing absolute values of the loadings. Second  (dotted-dashed line),  when ordering the cross-section by decreasing absolute values of the loadings. Third, when considering 100 random permutations of the cross-sectional units (dotted lines). In the latter case, we see that the evolution of the eigenvalue is always well described by a linear divergence in $n$. 

 \begin{figure}[t!]\label{Fig1}
 \begin{center}
\begin{tabular}{cc}
$\lambda_{X;1}^{(n)}/\lambda_{X;1}^{(n_0)}$ &$\lambda_{X;1}^{(n)}/\lambda_{X;1}^{(n_0)}$ \\
deterministic cross-sectional orderings&random cross-sectional orderings\\
 \includegraphics[width = .4\textwidth]{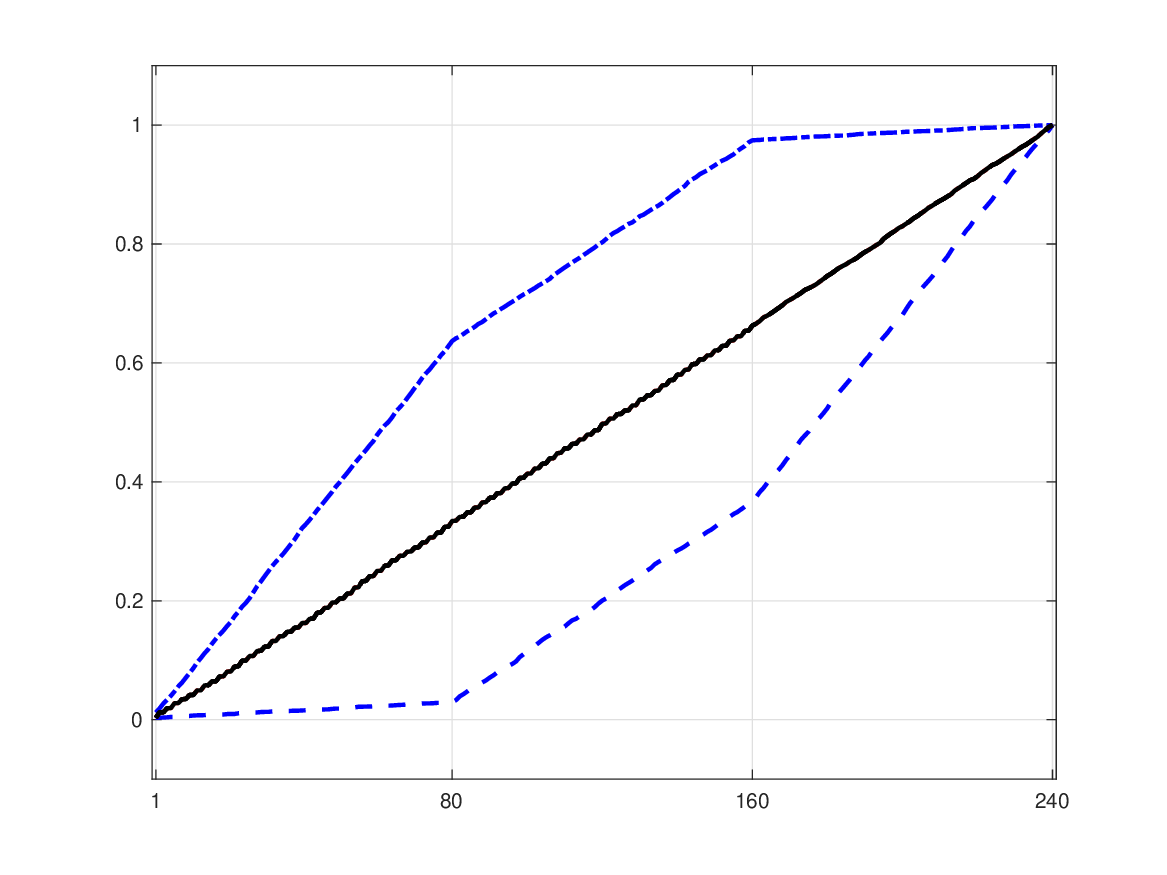}&
 \includegraphics[width = .4\textwidth]{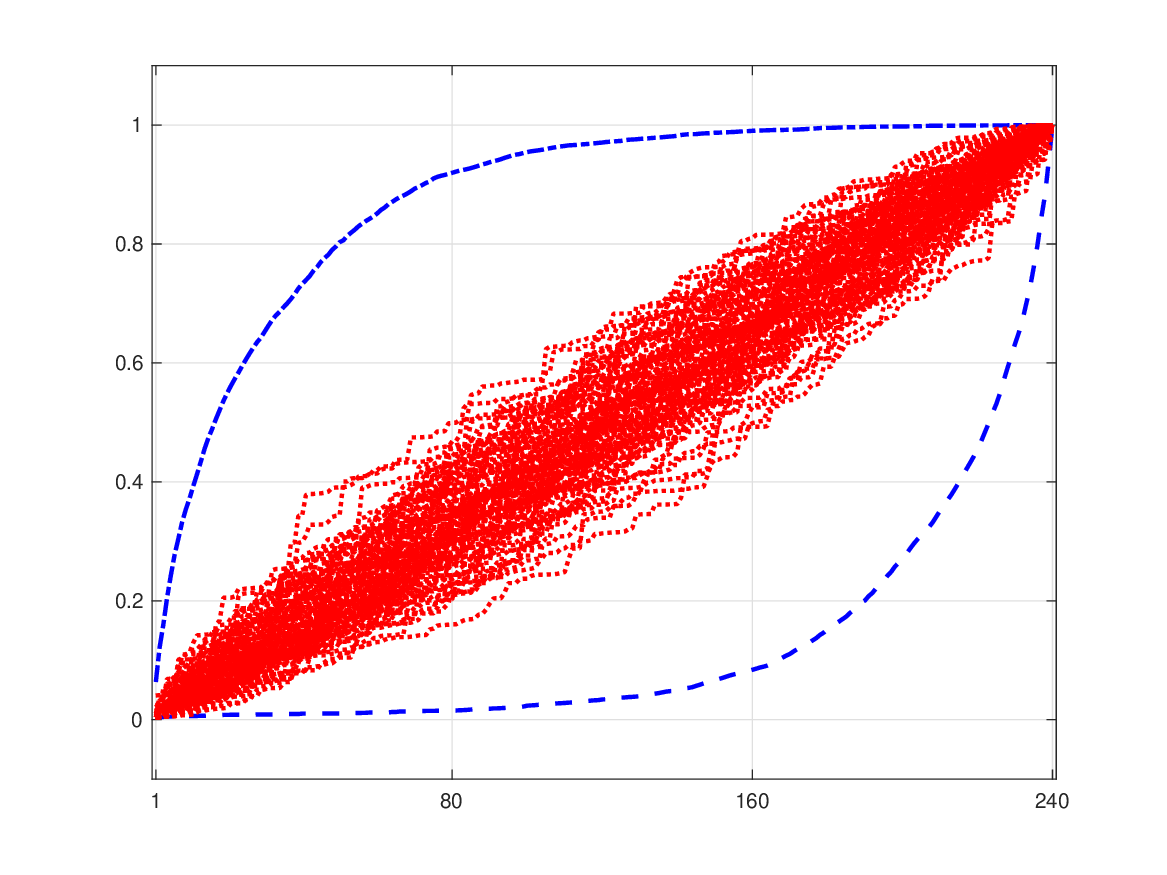}
 \end{tabular}
\caption{Evolution, for increasing $n=1,\ldots, n_0=240$, of the ratio  $\lambda_{X;1}^{(n)}/\lambda_{X;1}^{(n_0)}$), under different cross-sectional orderings.  Left panel (simulated data from \eqref{11}, $r=1$): increasing loadings (dashed line), decreasing loadings (dot-dashed line), and alternated small and large loadings (solid line). Right panel (simulated data from \eqref{11rand}, $r=1$): increasing loadings (dashed line), decreasing loadings (dot-dashed line), and~100 random cross-sectional permutations (dotted lines).}  \label{Fig1}
\end{center}
\end{figure}

 \begin{figure}[b!]\label{Fig2}
 \begin{center} 
 \begin{tabular}{cccc}
 $\lambda_{X;1}^{(n)}/\lambda_{X;1}^{(n_0)}$ &$\lambda_{X;2}^{(n)}/\lambda_{X;1}^{(n_0)}$ \\
 \includegraphics[width = .4\textwidth]{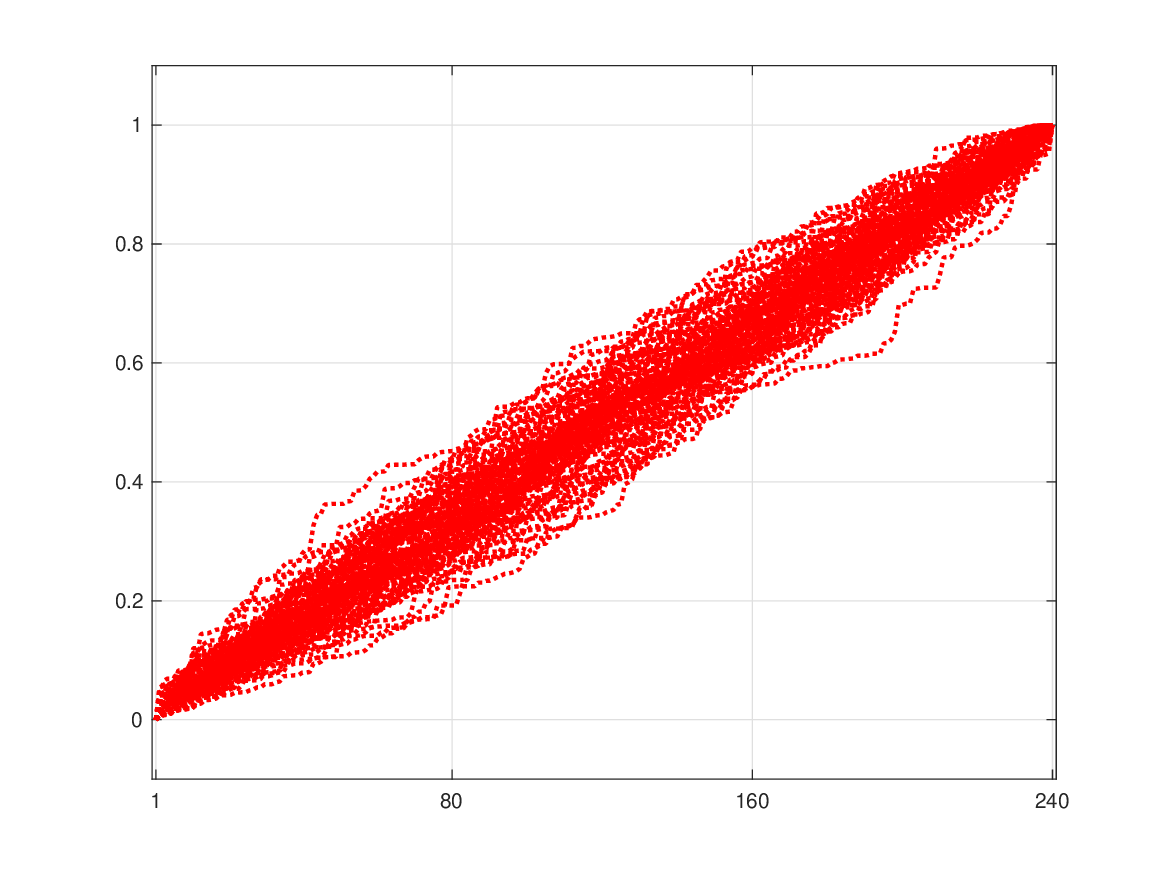}& \includegraphics[width = .4\textwidth]{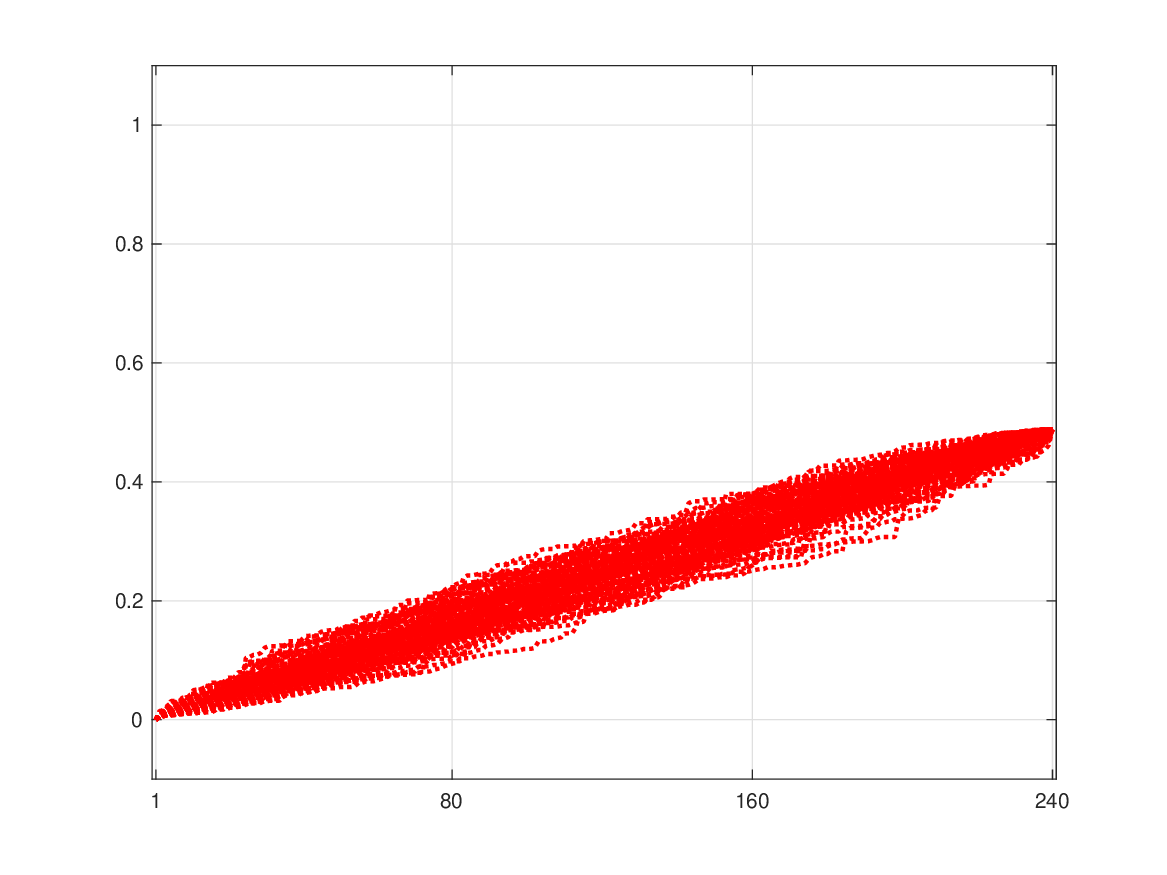}
 \end{tabular}
\caption{Simulated data from \eqref{11rand2}, $r=2$. Evolution for increasing $n=1,\ldots, n_0$ of the first  (left panel) and second (right panel) eigenvalues $\lambda_{X;1}^{(n)}$ and $\lambda_{X;2}^{(n)}$ (both normalized by $\lambda_{X;1}^{(n_0)}$), under 
100 random cross-sectional permutations (dotted lines).  }\label{Fig2}
\end{center}
\end{figure}
Third, we consider the same exercise but for a two-factor model
\begin{equation}\label{11rand2}
X_{it,m} = (B_{i1} u_{i1}) f_{1t}+(B_{i2} u_{i2}) f_{2t} + \varepsilon_{it}\qquad i=1,\ldots,n_0,\  t=1,\ldots,T,
\end{equation}
where $\{f_{1t}\vert\, t\in{\mathbb Z}\}$, $\{f_{2t}\vert\, t\in{\mathbb Z}\}$,  and  $\{\varepsilon_{it}\vert\, t\in{\mathbb Z}\}$, $i=1,\ldots,n_0$, 
 are $(n_0+2)$ mutually orthogonal white~noise processes, all with variance one, $B_{i1}=B_{i2}$, $i=1,\ldots, n_0$, are  generated as $B_i$ in \eqref{11},  with~$u_{i1}\sim N(0,1)$, and 
 $u_{i2}\sim N(0,0.5)$, $i=1,\ldots, n_0$. 
 In Figure~\ref{Fig2} we show  the evolution for increasing~$n=1,\ldots, n_0$ of the first and second eigenvalues $\lambda_{X;1}^{(n)}$ (left panel) and $\lambda_{X;2}^{(n)}$  (right panel) (both normalized by $\lambda_{X;1}^{(n_0)}$) of~${\boldsymbol\Gamma}^{(n)}_{X}$ 
 when considering 100 random permutations of the cross-sectional units. We see that the evolution of both eigenvalues is always well described by a linear divergence in $n$, despite having different slopes due to the fact that eigenvalues are distinct.

\subsection{Factors and cross-sectional permutations: exchangeability}\label{Sec62}

{Distributional symmetries and group invariance properties are among the most convincing  and forceful  arguments giving  statistical inference problems   a structure which, when it exists,  dramatically   impacts their solutions. Invariance arguments have a long history in mathematical statistics (see Chapter~6 in \cite{lehmann2006testing}); their pertinence, in a broad variety of problems, has been highlighted recently in    \citet{austern2022limit}. }

{Stationarity in time is one of those fundamental invariance properties and, under our fully nonparametric approach (no parameters), is the only assumption we are imposing in order to obtain a factor model decomposition. Namely, we assume} that the observed  panel is the finite $(n\times T)$ realization of a stochastic process $\bf X$ of the form~$\big\{X_{it} \vert  \, i\in{\mathbb N}, t\in{\mathbb Z}
\big\}$ in the class $\mathcal C$ of real-valued  second-order time-stationary processes indexed on~${\mathbb N} \times{\mathbb Z}$. Second-order stationarity
  imposes a restriction (stationarity) on  time dependence, namely,  
  \begin{ass}\label{exchdef}
{\rm 
The stochastic process $\big\{X_{it} \vert i\in{\mathbb N}, t\in{\mathbb Z}
\big\}$ is  { \it second-order stationary}---namely,  for any~$i,\, i\pr$ in~$\mathbb N$, any $t,\, t\pr\in\mathbb Z$, and any   lag $h\in{\mathbb Z}$, 
 $ {\rm E}\left[X_{i  t}X_{i\pr t\pr}
\right] = {\rm E}\left[X_{i, t+h}X_{i\pr , t\pr +h}
\right] $.
}\end{ass}

This, however,  does not say anything about cross-sectional dependence and, in particular,   does not reflect the   irrelevance of cross-sectional ordering. If that  irrelevance is to be incorporated into the assumptions, the class $\mathcal C$ has to be restricted further {by imposing   {\it  (second-order) cross-sectional exchangeability}---another  strong structural invariance property (of the same nature as (second-order) time-stationarity) discussed in \citet{austern2022limit}; that  assumption of exchangeability   has been used in \cite{Barigozzietal24}    to derive very complete  asymptotic distributional results for the GDFM estimators of  Forni et al.~(2015, 2017).}


\begin{ass}\label{exchdef}
{\rm 
The stochastic process $\big\{X_{it} \vert i\in{\mathbb N}, t\in{\mathbb Z}
\big\}$ is  { \it (second-order) cross-sectionally exchan\-geable}---that is, for any $k\in{\mathbb N}$, any $k$-tuple $i_1,\ldots,i_k$, and any permutation $\pi$ in the set $\Pi_k$ of the permutations of the integers~$(1,\ldots,k)$, the distributions (the  second-order moments) of the  $k$-dimensional stochastic subprocesses 
$$\big\{(X_{i_1t},\ldots,X_{i_kt}) \vert \,   t\in{\mathbb Z}\big\}
\quad\text{and}\quad
 \big\{(X_{i_{{  \pi(1)}}t},\ldots , X_{i_{{  \pi(k)}}t}
 ) \vert \,    t\in{\mathbb Z}
\big\}
$$
are the same. 
}\end{ass}

Clearly, exchangeability implies second-order exchangeability, viz.  $ {\rm E}[X_{i_{\ell} t}X_{i_{m} t\pr}
] = {\rm E}[X_{i_{\pi(\ell )}t}X_{i_{\pi(m)}t\pr}
] $ for any $\ell$ and $m$ in $(1,\ldots,k)$,  any $t$ and $t\pr$ in~$\mathbb Z$, and any permutation $\pi\in\Pi_k$. While second-order exchangeability, just as second-order stationarity, is mathematically sufficient for our needs, ``full'' exchangeability and ``full'' stationarity are the most natural and intuitively justifiable assumptions here. Denote by ${\mathcal C}_*\subset\mathcal C$ the class of second-order stationary  second-order exchangeable processes~$\mbf X=\big\{X_{it} \vert\, i\in{\mathbb N}, t\in{\mathbb Z}
\big\}$.

 Let us show that, under exchangeability, eigenvalue divergence only can be linear in $n$. 

The process  ${\bf X}=\big\{X_{it} \vert i\in{\mathbb N}, t\in{\mathbb Z}
\big\}\in{\mathcal C}_*$ can be rewritten as ${\bf X}=\big\{\{ X_{it}\vert i\in{\mathbb N}\}\vert t\in{\mathbb Z}\big\}\in{\mathcal C}_*$, emphasizing the fact that it can be considered as an infinite-dimensional time series.  Its distribution then is naturally  described as resulting from a  
%
  two-step random mechanism: (Step~I) the stochastic selection, via some unspecified distribution~${\mathbb P}$, of the distributional features ${\rm P}_{\bf X}$, say, 
 of the infinite-dimensional   time series $\big\{\{ X_{it}\vert i\in{\mathbb N}\}\vert t\in{\mathbb Z}\big\}$,  
followed by~(Step~II) a  realization over time  of the same. 

A more formal description of Step~I can be given as follows. Consider the space 
$${\mathcal F}\coloneqq \Big\{{\mathrm P}_{\bf X}: {\mathrm P}_{\bf X}\text{ the probability distribution of a stochastic process ${\bf X}\in\mathcal C_*$}
\Big\}
$$
 equipped  with the $\sigma$-field ${\mathcal A}$ of all Borel (with respect to the topology of weak convergence) sets.  Denoting by ${\mathbb P}$ a  distribution over some adequate probability space $(\Omega, {\mathcal A}_\Omega)$, let $\omega\mapsto{\rm P}_{\bf X}(\omega)$ denote a measurable map from $(\Omega, {\mathcal A}_\Omega)$ to $({\mathcal F} ,{\mathcal A})$: ${\rm P}_{\bf X}$ then is the  random   distribution of an infinite-dimensional   time series,   
  the   $n\times n$ spectral density matrices ${\boldsymbol\Sigma}\n_{\phantom{0}}(\theta)$ and   $n\times n$ covariance matrices ${\boldsymbol\Gamma}\n$  of which are   random as well, and so are random, as well as also their eigenvalues $\lambda\n_j(\theta)$ and $\lambda\n_{X;j}$.

The distribution  ${\mathbb P}$ here plays the role of a nuisance. Statistical practice in such cases consists in conducting inference on the realization observed in Step~II conditional on the (unobserved) result of Step~I (see, e.g.,  Chapter 10 of \citet{lehmann2006testing}), so  that ${\mathbb P}$  needs no further description. Under such conditional approach, the random distribution
 ${\rm P}_{\bf X}$  of the  process ${\bf X}\in{\mathcal C}_*$  
 of which the observed panel   is a finite realization is  treated as unknown but fixed. Hence  ${\boldsymbol\Sigma}\n({\theta})$,   ${\boldsymbol\Gamma}^{(n)}$, and their eigenvalues are random variables  treated as unknown but fixed quantities on which conditions such as~\eqref{7} or \eqref{9} are imposed: under unconditional form, these conditions actually should be imposed~$\mathbb P$-almost surely. 
 


Now, for simplicity, let us consider the first eigenvalue $\lambda^{(n)}_1$ of the   $n\times n$   (conditional  on the result of Step~I)  covariance matrix  ${\boldsymbol\Gamma}^{(n)}$. Let $\delta^{({\nu})}_1\coloneqq \lambda^{({\nu})}_1- \lambda^{({\nu}-1)}_1$, ${\nu}\geq 2$ denote  the contribution of cross-sectional item ${\nu}$ to $\lambda^{(n)}_1$. Then, $\lambda^{(n)}_1 = \lambda ^{(1)}_1 + \sum_{{\nu}=2}^n\delta^{({\nu})}_1$. 
Denote by $\mathbb E$ the expectation associated with $\mathbb P$:    due to exchangeability, ${\mathbb E}[\delta^{({\nu})}_1] \eqqcolon \delta _1$ does not depend on $\nu$. Hence, 
 ${\mathbb E}[\lambda^{(n)}_1] = {\mathbb E}[ \lambda ^{(1)}_1] +  (n-1)\delta_1$. Since $\delta^{({\nu})}_1$ is a nonnegative variable, two cases are possible. Either $\delta _1=0$ $\mathbb P$-a.s., and $\lambda^{(n)}_1 = \lambda^{(1)}_1$ a.s.\ is  $\mathbb P$-a.s.\ bounded as $n\to\infty$; or $\delta _1> 0$ and ${\mathbb E}[\lambda^{(n)}_1]$ explodes at rate $n$, which is incompatible with $\lambda^{(n)}_1/n^{\alpha} \to c$  $\mathbb P$-a.s.\  with~$c<\infty$ for $\alpha <1$ and $c>0$ for $\alpha >1$.    
This precludes non-linear growth of ${\mathbb E}[\lambda^{(n)}_1]$.   A similar reasoning holds for $\lambda^{(n)}_j$, $j>1$,  as well as for the  dynamic eigenvalues $\lambda_j^{(n)}(\theta)$.

If the irrelevance of the cross-sectional ordering is taken into account, thus, rate-weak factors {\it are nowhere}---much ado about nothing!

 \begin{figure}[b!]\label{Fig3}
 \begin{center} 
 \begingroup

\setlength{\tabcolsep}{-7pt}
 \begin{tabular}{cccc}
 \hspace{-0mm} $\scriptstyle\lambda_{X;1}^{(n)}/\lambda_{X;1}^{(n_0)}$ \hspace{-0mm} &$\scriptstyle\lambda_{X;2}^{(n)}/\lambda_{X;1}^{(n_0)}$  \hspace{-0mm}& $\scriptstyle\lambda_{X;3}^{(n)}/\lambda_{X;1}^{(n_0)}$  \hspace{-0mm}&$\scriptstyle\lambda_{X;4}^{(n)}/\lambda_{X;1}^{(n_0)}$ \\
 \hspace{-0mm}  \includegraphics[width = .27\textwidth]{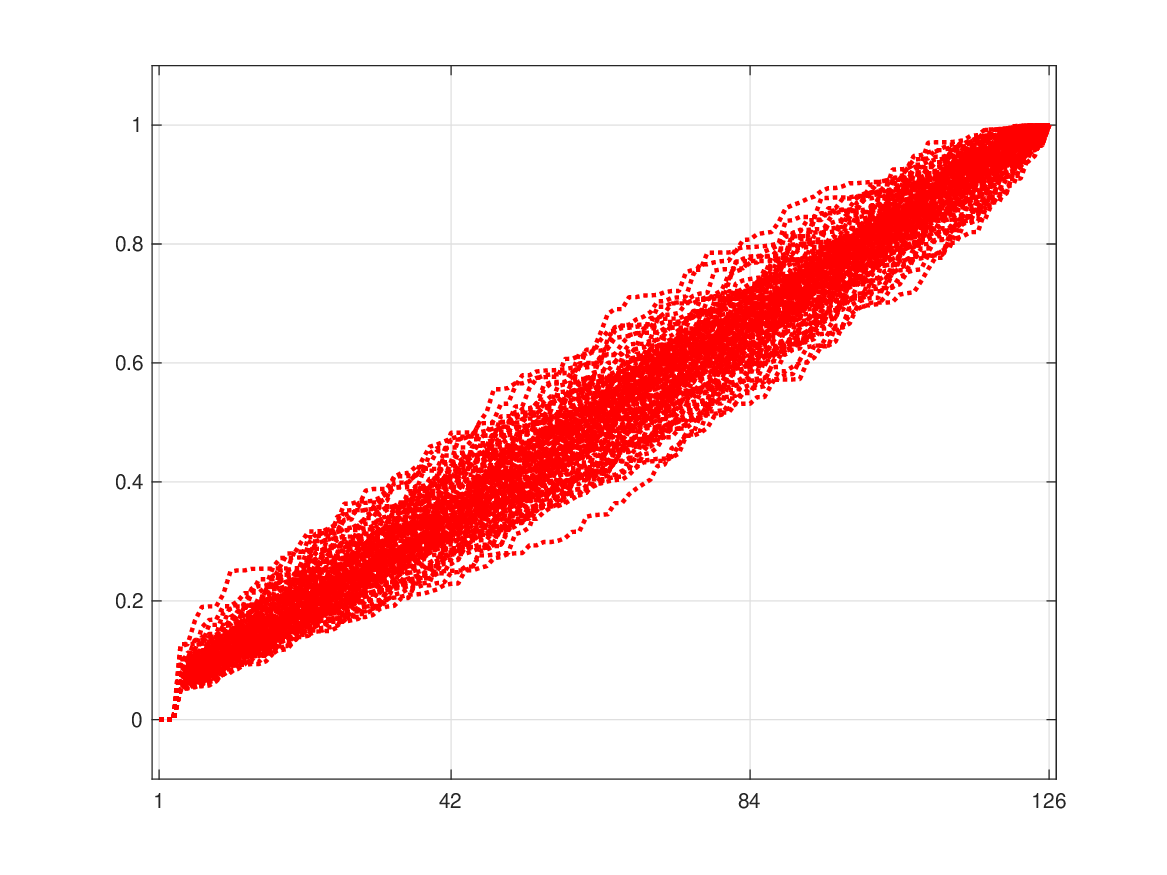} \hspace{-0mm}& \includegraphics[width = .27\textwidth]{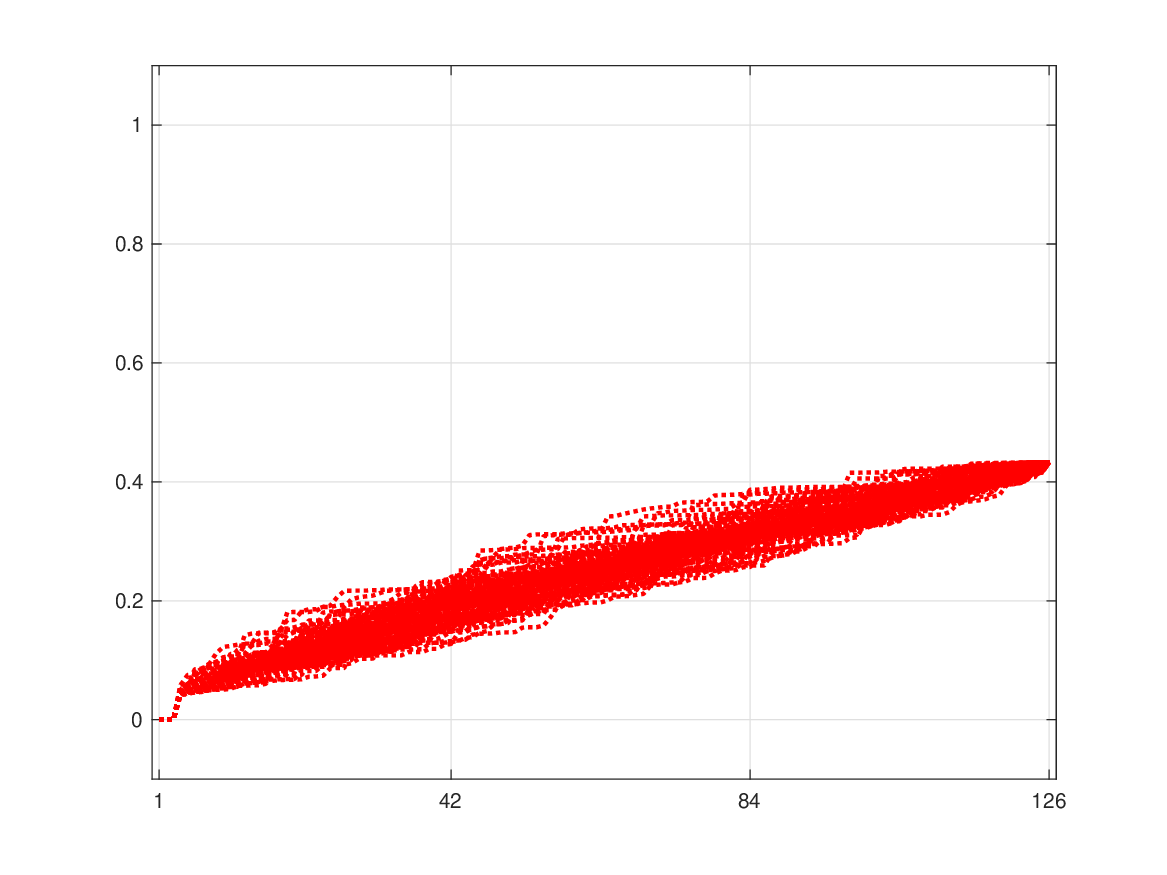} \hspace{-0mm}& 
 \includegraphics[width = .27\textwidth]{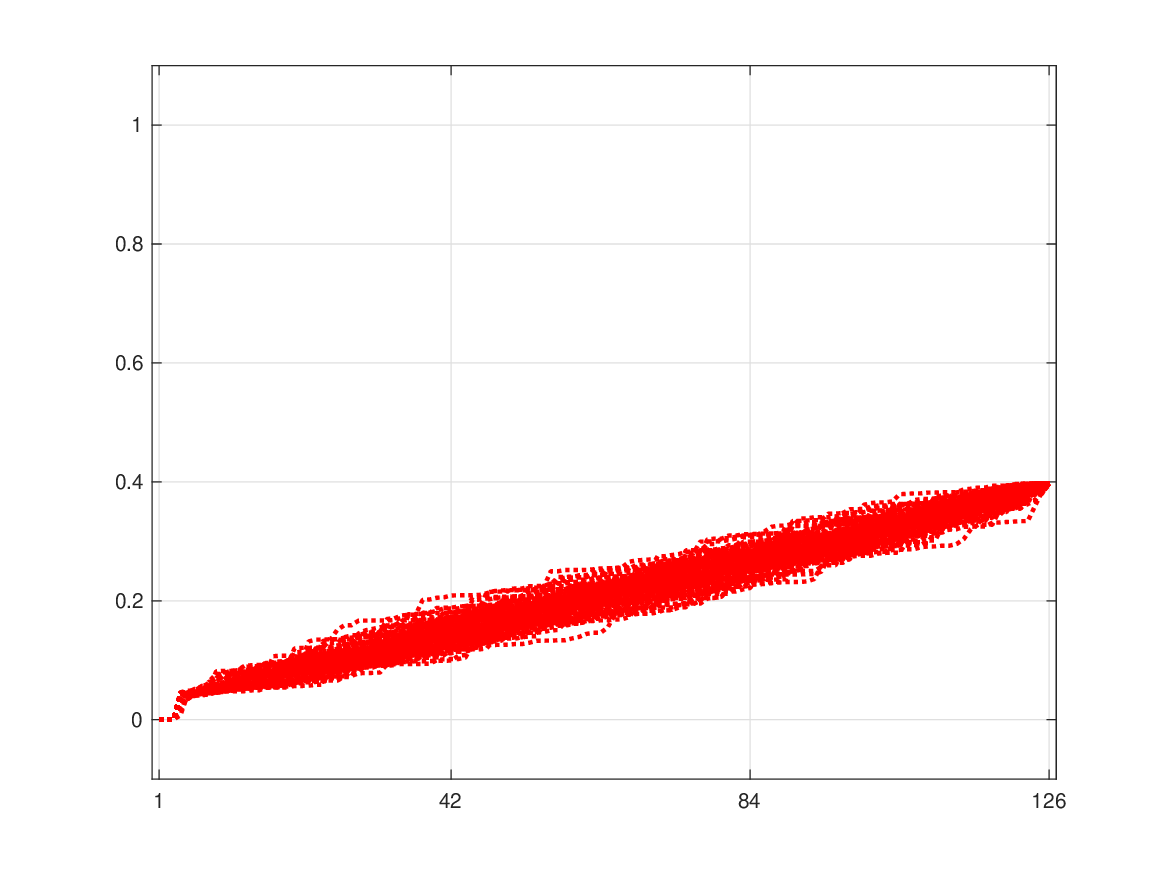} \hspace{-0mm}& \includegraphics[width = .27\textwidth]{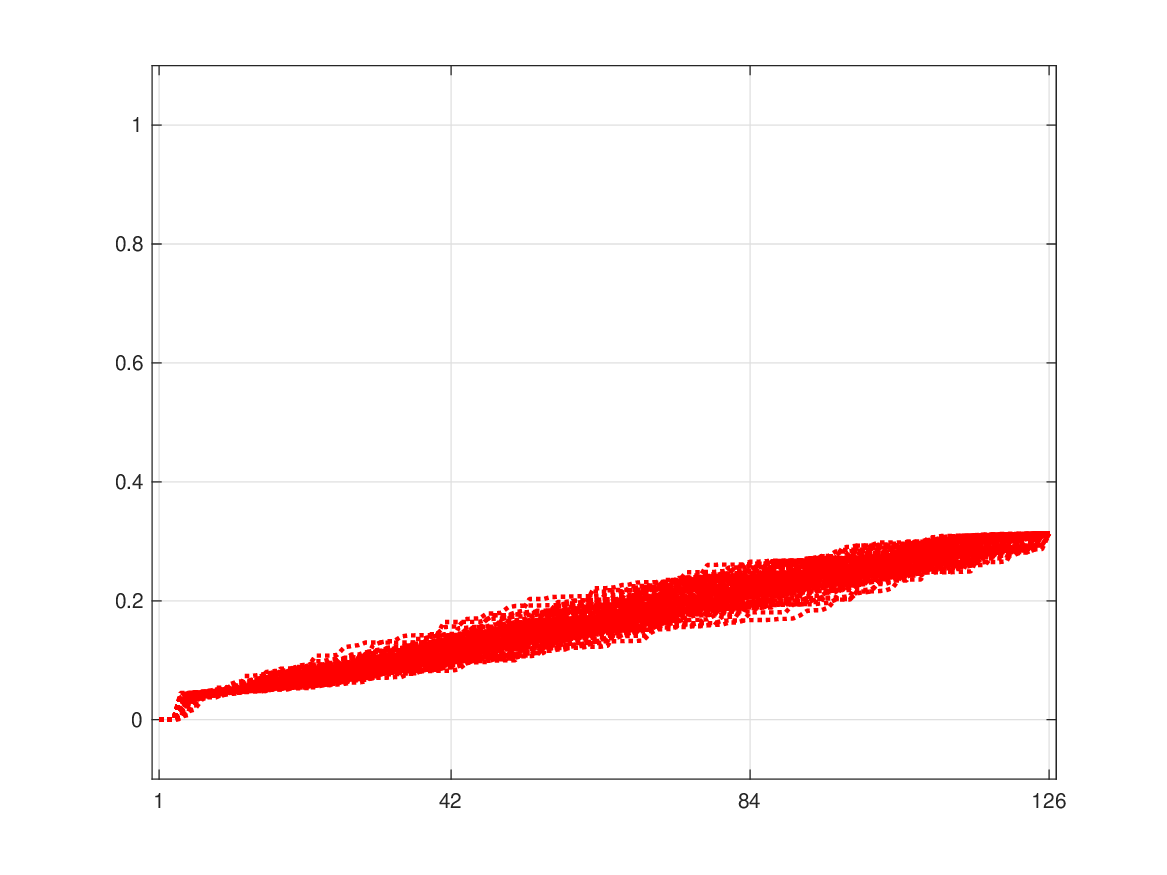}\\
 \hspace{-0mm}  $\scriptstyle\lambda_{X;1}^{(n)}/\lambda_{X;5}^{(n_0)}$  \hspace{-0mm}&  \hspace{-0mm} $\scriptstyle\lambda_{X;6}^{(n)}/\lambda_{X;1}^{(n_0)}$  \hspace{-0mm}& $\scriptstyle\lambda_{X;7}^{(n)}/\lambda_{X;1}^{(n_0)}$  \hspace{-0mm}&$\scriptstyle\lambda_{X;8}^{(n)}/\lambda_{X;1}^{(n_0)}$ \\
 \hspace{-0mm} \includegraphics[width = .27\textwidth]{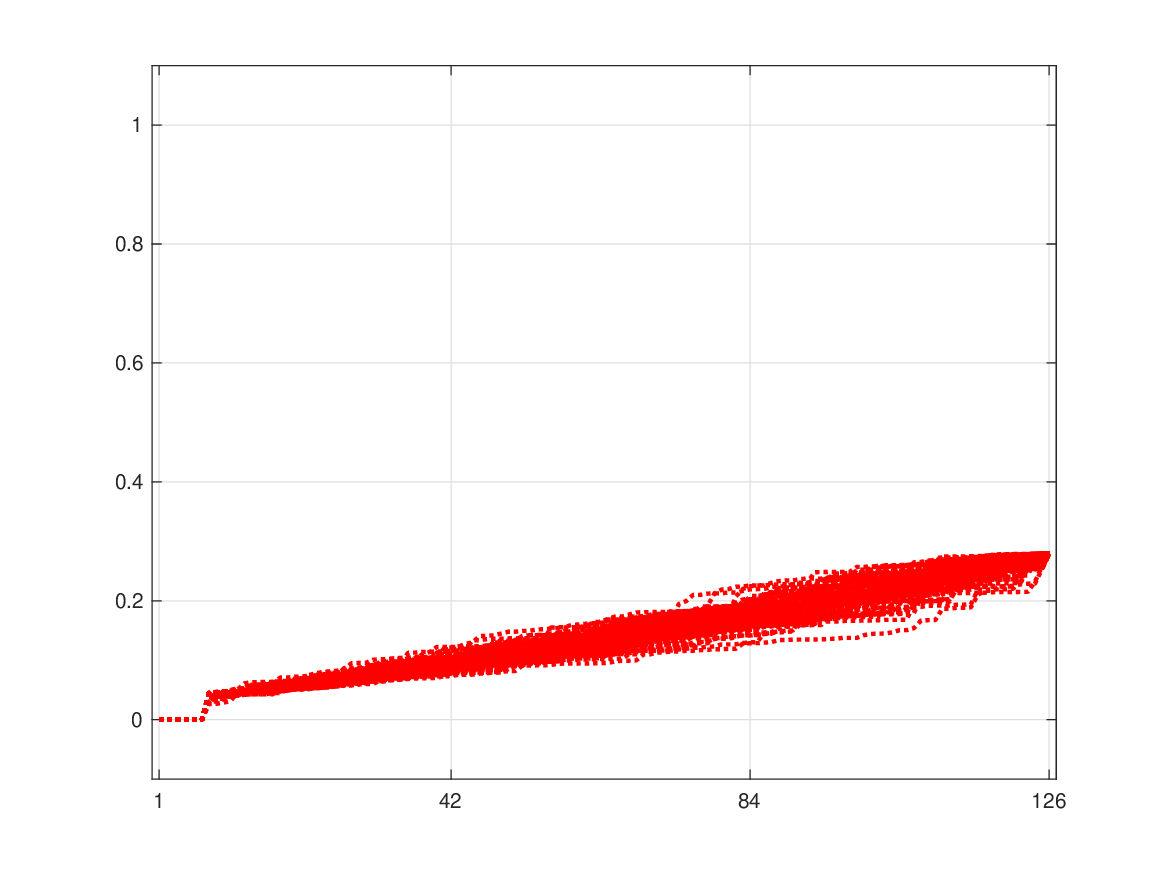} \hspace{-0mm}& \includegraphics[width = .27\textwidth]{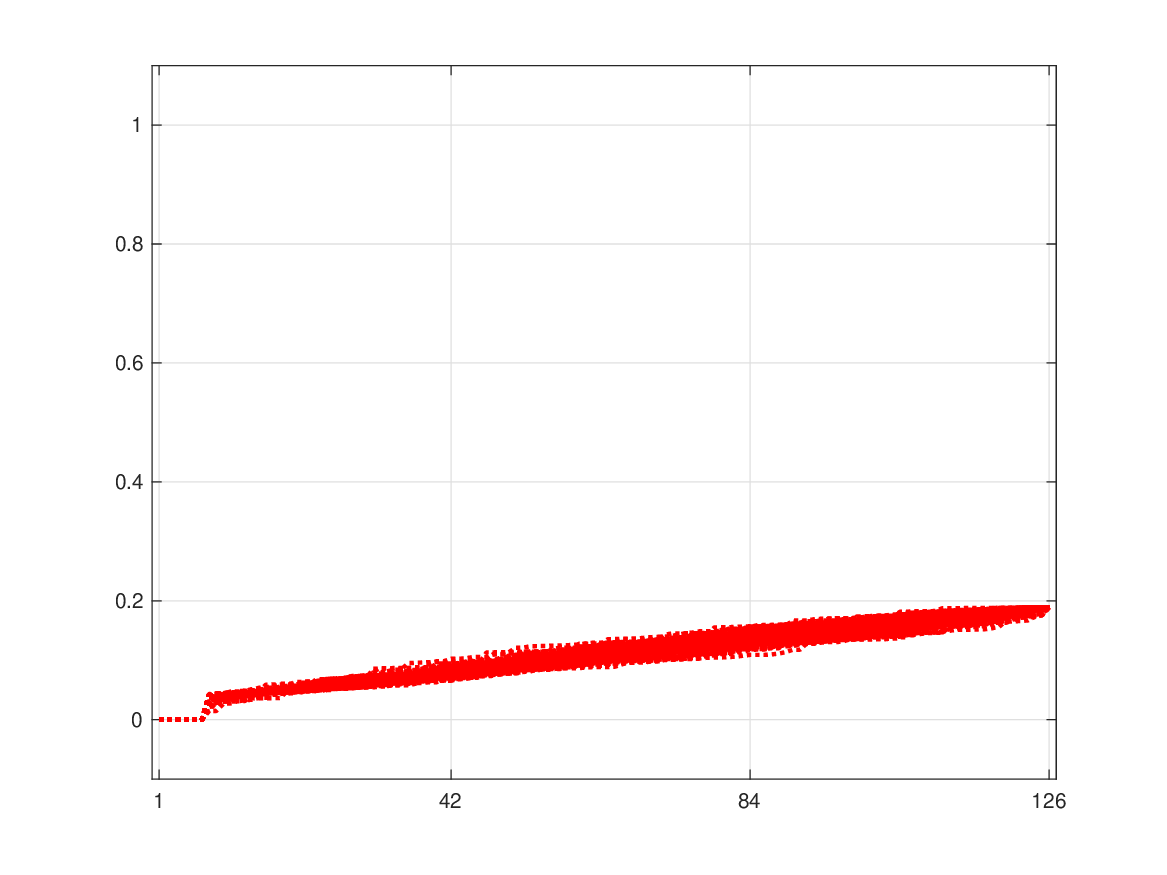} \hspace{-0mm} & 
 \includegraphics[width = .27\textwidth]{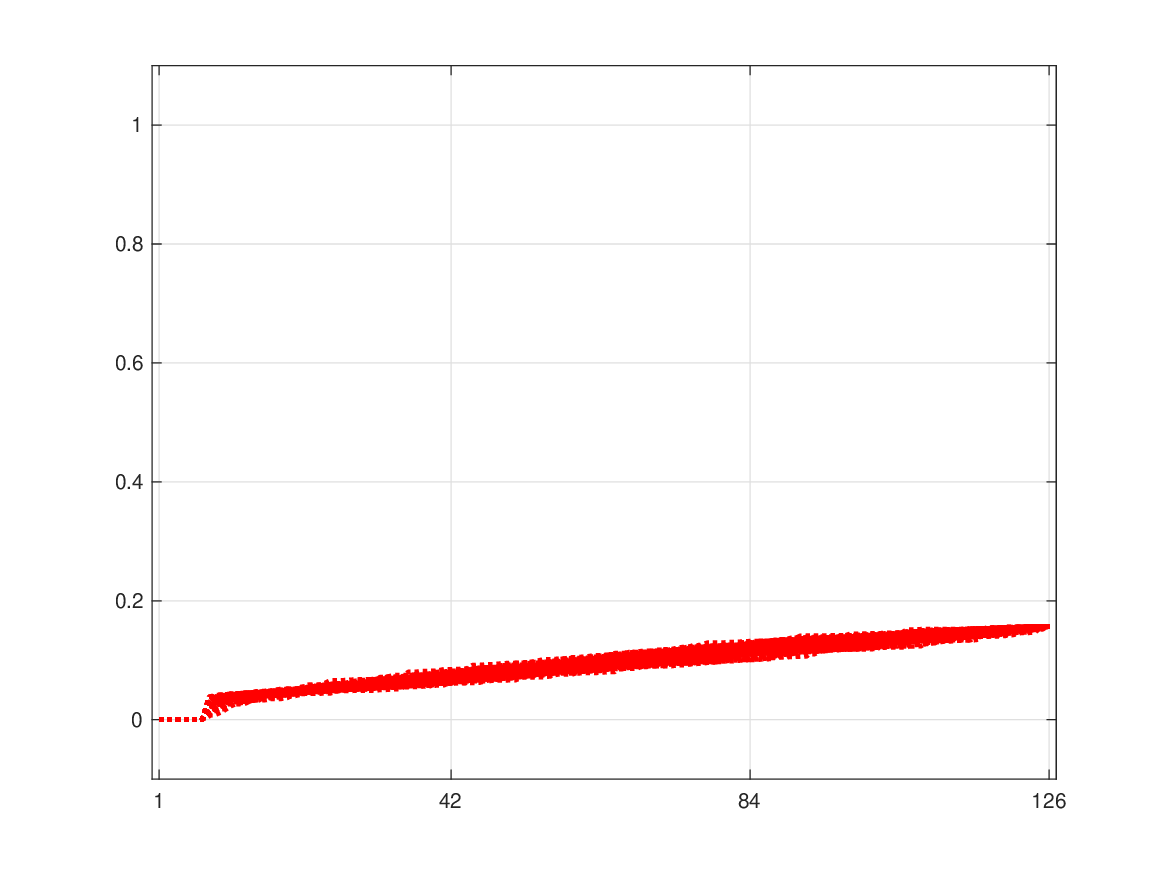} \hspace{-0mm}& \includegraphics[width = .27\textwidth]{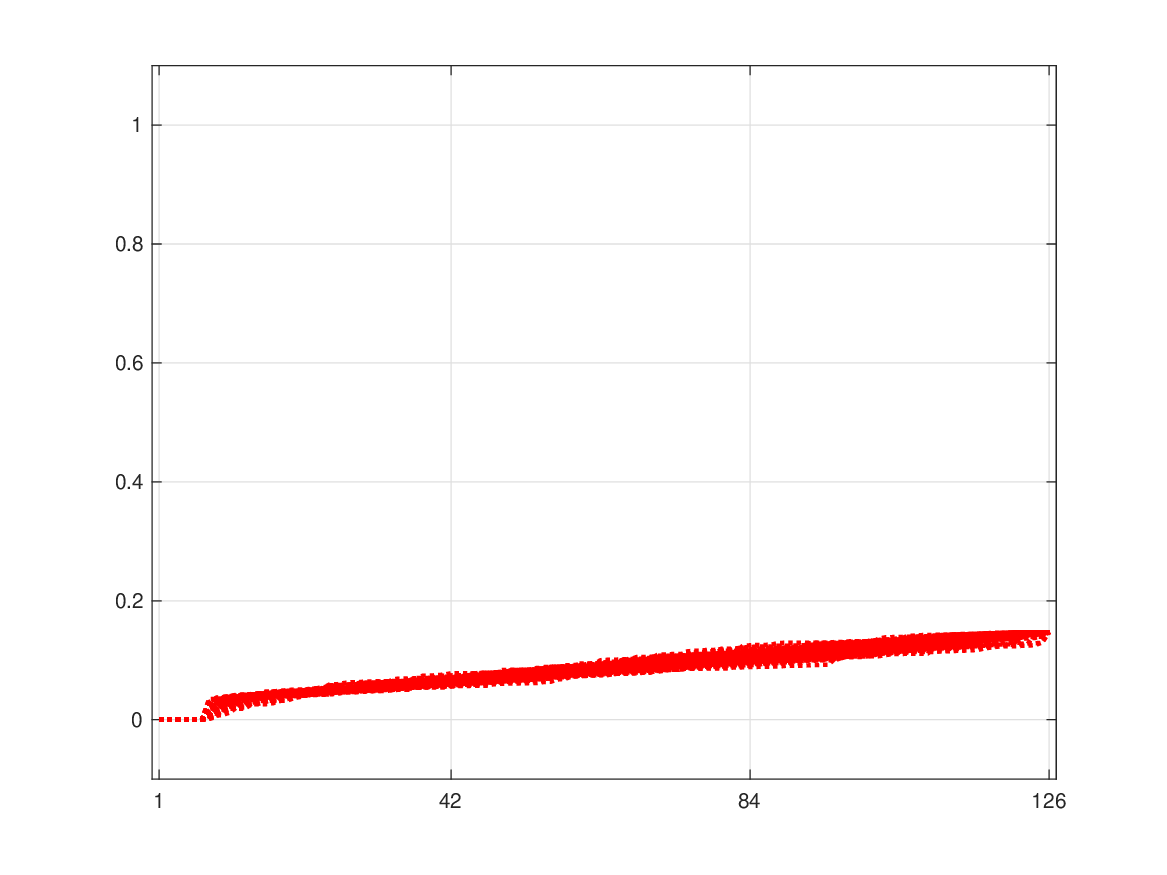}
  \end{tabular}
  \endgroup
\caption{FRED-MD dataset  \citep{mccracken2016fred}, $r=8$. Evolution, for increasing $n=1,\ldots, n_0$, of the eigenvalues $\lambda_{X;k}^{(n)}$, $k=1,\ldots, 8$ (all normalized by $\lambda_{X;1}^{(n_0)}$) under 100 random cross-sectional permutations.\vspace{-3mm}}\label{Fig3}
\end{center}
\end{figure}
In Figure~\ref{Fig3}, as a numerical illustration of that fact, we consider the FRED-MD dataset  described by \citet{mccracken2016fred}, which consists of~$n_0=126$ monthly times series covering both the real and nominal sectors of the U.S.~economy and also  includes   labor market indicators and financial variables (see \url{https://research.stlouisfed.org/econ/mccracken/fred-databases/}). The data is transformed to stationarity and missing values are imputed by means of the routines made available by \citet{mccracken2016fred}, which
produces a balanced panel, with a sample running from April 1959 through March 2024---an observation period of~$T=780$ time points. 

The cross-sectional ordering, in this dataset, is fully arbitrary, hence should be meaning\-less. Standard methods  \citep{BaiNg02} for determining the number of static factors point to as much as $\wh r=8$ factors.    Figure~\ref{Fig3}   shows  the evolution, for increasing $n=1,\ldots, n_0$, of the eigen\-values~$\lambda_{X;k}^{(n)}$, $k=1,\ldots, 8$ (all normalized by $\lambda_{X;1}^{(n_0)}$) of~${\boldsymbol\Gamma}^{(n)}_{X}$ when considering 100 random permutations of the cross-section. The plots show  that the evolution of all eigenvalues is   well described by a linear (in $n$) divergence, with decreasing (in $k$) slopes. These slopes, for the two or three last eigenvalues, are small  and one may be tempted to consider them as rate-weak and speculate about the value $\alpha$ in some  sublinear divergence rate $n^{\alpha}$: this is, however, disputable and incompatible with the irrelevance of cross-sectional ordering.

\color{black}
\section{Identifying the number of factors}\label{Sec7} 

From the above discussions, it is clear that  the numbers of factors ($r$ in the static decomposition~\eqref{8}\linebreak  satisfying~\eqref{9} and $q$  in the GDFM decomposition \eqref{6} satisfying~\eqref{7}) are only asymptotically  identified as $n\to\infty$, and coincide with the number of diverging eigenvalues of the covariance and spectral density matrices, respectively. In practice, with a finite-$(n,T)$ realization $\{X_{it} \vert\, i=1,\ldots,n ,  \  t=1,\ldots, T\}$ of the  process~$\{X_{it} \vert\, i\in~\!\mathbb{N},\,~\!t\in~\!\mathbb{Z}\}$, an estimation of these  numbers has to be based on estimators, $\wh{\bm\Gamma}^{(n)}_X$ or $\wh{\bm\Sigma}^{(n)}_X(\theta)$, of these matrices  and the corresponding empirical eigenvalues $\wh{\lambda}^{(n)}_{X;j}$ or $\wh{\lambda}^{(n)}_{X;j}(\theta)$,  $\theta\in[-\pi,\pi]$. 

Whether the objective is the  identification/estimation  of $r$ or $q$, two main strategies can be found in the literature.
The most popular one---call it the {\it eigengap detection} strategy---is based on the fact that~$r$ and $q$ (unless they are zero) are characterized by an arbitrarily large ``eigengap'' between  unbounded and bounded eigenvalues. More precisely, $r$ is the largest value of $j$ such that  ${\lambda}^{(n)}_{X;j} - {\lambda}^{(n)}_{X;j+1}\to\infty$   as $n\to\infty$, 
 and $q$  the largest value of $j$ such that  {$\int_{-\pi}^\pi \l\{{\lambda}^{(n)}_{X;j}(\theta)- {\lambda}^{(n)}_{X;j+1}(\theta)\r\}{\rm d}\theta\to\infty$ }  as $n\to\infty$. A sensible strategy, thus,    consists in selecting  $\widehat{r}^{\,(n,T)}$ and $\widehat{q}^{\,(n,T)}$ as the maximal $j$'s yielding a  ``large eigengap'' between the empirical counterparts 
 $ \wh{\lambda}^{(n)}_{X;j} -  \wh{\lambda}^{(n)}_{X;j+1}$ and 
 { $\int_{-\pi}^\pi \l\{\wh{\lambda}^{(n)}_{X;j}(\theta)- \wh{\lambda}^{(n)}_{X;j+1}(\theta)\r\}{\rm d}\theta$ }of these quantities (note that, as functions of the frequency $\theta\in[-\pi,\pi]$, these~$\wh{\lambda}^{(n)}_{X;j}(\theta)$ do not cross). Depending on what is meant with a ``large eigengap,'' several implementations of this strategy have been proposed. The prototype  for the identification of $r$ is the information criterion method of \cite{BaiNg02}, which is also the most popular, followed by many others among which  \cite{onatski2010determining}, based on  the asymptotic distribution of the $\wh{\lambda}^{(n)}_{X;j}$'s, and the  eigenvalue ratio approaches  by 
 \citet{ahn2013eigenvalue} and   \citet{trapani2018randomized}---to quote only a few.

A different strategy  is based on the fact that the divergence of eigenvalues is a property of sequences of the type ${\lambda}^{(m)}_{X;j}, {\lambda}^{(m +1)}_{X;j}, \ldots $ or ${\lambda}^{(m)}_{X;j}(\theta), {\lambda}^{(m +1)}_{X;j}(\theta), \ldots $ This strategy therefore considers the corres\-pon\-ding empirical   trajectories  rather than restricting to  the ``terminal'' empirical values~$\wh{\lambda}^{(n)}_{X;j}$ and~$\wh{\lambda}^{(n)}_{X;j}(\theta)$ and was first proposed by \citet{HallinLiska07} for the identification of $q$ in the GDFM.   That method is designed to minimize, over a grid of increasing values of the cross-sectional dimension, the trace of the idiosyncratic spectral density subject to a penalty which is tuned by means of a positive\linebreak  constant~$c\in~\![0,c_{\max}]$   where $c_{\max}>0$ is some pre-specified maximal value.  More precisey, for  a  grid~${\mathfrak G}\n\coloneqq \{0=m_1<\ldots< m_\nu <\ldots <m_{N(n)}=n\}$ of  increasing cross-sectional dimensions and  a~grid~${\mathfrak C}\coloneqq \{0<c_1<\ldots <c_\ell <\ldots c_L =c_{\max}\}$ of increasing tuning constants, this method considers, for each~$m\in {\mathfrak G}\n$, fixed $n$, and  $c\in{\mathfrak C}$, the minimization problem
\begin{align}
&\min_{1\le k\le q_{\max}} \text{IC}^{(m)}(k;c),\;\text{ where }\; \text{IC}^{(m)}(k;c)\coloneqq  
\sum_{j=k+1}^{m} \int_{-\pi}^\pi \wh{\lambda}^{(m)}_{X;j}(\theta) \mathrm d\theta 
+k c\; \mathrm p(m)
 \nn
\end{align}
for some    
  penalty function $\mathrm p$ and pre-specified 
 $q_{\max}$. 
 Solving   these  problems yields, for each   $c\in {\mathfrak C}$,  a sequence $\check{q}_c\n(m)$, $m\in~{\mathfrak G}\n$ of possible numbers of factors.  
 For adequate penalty function $\mathrm p$, it can be shown that all these sequences, irrespective of $c$,  are consistent as $n\to\infty$.  The choice of an ``optimal penalty''~$c$,   denote it as $c^*$, can be based on the fact that overpenalization ($c$ too large) yields consistency from below while underpenalization yields consistency from above; this suggests selecting~$c^*$ as any value of~$c$ achieving, as $m$ ranges over ${\mathfrak G}\n$,  the most ``stable'' trajectory of $\check{q}_c\n(m)$ values  (typically, an interval of $[0, c_{\max}$). The selected number of factors then is~$\widehat{q}^{\,(n,T)}\coloneqq \check{q}_{c^*}\n(n)$.  \citet{HallinLiska07} establish the consistency consistency under linearly divergent eigenvalues;   \cite{Ona24} shows that this consistency still holds when the factors are extremely (rate-)weak.
 
An analogue of this method for the static case and the estimation of $r$  is proposed by \citet{ABC10}, who refine the information criterion approach of \citet{BaiNg02} by introducing, as in \citet{HallinLiska07},  a   tuning of the penalty which enhances its finite-sample performance.

The   performance of these methods, however,  crucially depends on the following two issues: 
\ben
\item[(a)] the quality of $\wh{\lambda}^{(n)}_{X;j}$ and $\wh{\lambda}^{(n)}_{X;j}(\theta)$ as  estimators of   ${\lambda}^{(n)}_{X;j}$ and ${\lambda}^{(n)}_{X;j}(\theta)$, and 
\item[(b)] the quality of  ${\lambda}^{(n)}_{X;j}$ and~${\lambda}^{(n)}_{X;j}(\theta)$ as   approximations of the corresponding eigenvalues  ${\lambda}_{X;j}$ and~${\lambda}_{X;j}(\theta)$ of the infinite-dimensional covariance and spectral density matrices  ${\bm\Sigma}_X$ and ${\bm\Sigma}_X(\theta)$ of the  observed process~$\{X_{it} \vert\, i\in\mathbb{N} ,  \  t\in\mathbb{Z}\}$.
\een
The former (a) is an estimation problem, the latter a population feature on which we have no control.  
As far as (a) is concerned,  many results   on the estimation of large covariance ands spectral density matrices (and their eigenvalues) are available in the literature, and    we will not discuss this here---see, e.g., \citet{wu2018asymptotic} and \citet{zhang2021convergence} for a modern treatment. 


A third point, moreover, is to be taken into account in view of the irrelevance of cross-sectional ordering and the arguments developed in Section~\ref{sec5}:
\ben
\item[(c)] the 
 selected values $\widehat{q}^{\,(n,T)}$ and $\widehat{r}^{\,(n,T)}$ should be invariant under cross-sectional permutations.
\een 
All  ``eigengap detection'' methods  naturally satisfy this invariance requirement since $\wh{\lambda}^{(n)}_{X;j}$ and $\wh{\lambda}^{(n)}_{X;j}(\theta)$ do for all~$j$.   But  the \citet{HallinLiska07} and  \citet{ABC10} methods do not because the se\-quences~${\lambda}^{(m)}_{X;j}(\theta)$, $m=j,\ldots, n$ very much depend on the cross-sectional ordering. A natural way to eliminate this dependence in the \citet{HallinLiska07} method consists in averaging the $\wh{\lambda}^{(m)}_{X;j}(\theta)$'s over several random cross-sectional permutations. That average typically stabilizes very rapidly  with the number of permutations 
 and, when performed on the resulting sequences~$\overline{\lambda}^{(m)}_{X;j}(\theta)$, $m=j,\ldots, n $, the  \citet{HallinLiska07}  method satisfies (c). The same remark holds, and the same averaging is in order for the estimator $\widehat{r}^{\,(n,T)}$ of $r$ based on the \citet{ABC10} method and the static eigenvalues $\wh{\lambda}^{(n)}_{X;j}$.

The literature is mostly  focused on the consistency properties, as~$(n,T)\to\infty$,  of $\widehat{q}^{\,(n,T)}$ as an estimator of~$q$ (of $\widehat{r}^{\,(n,T)}$ as an estimator of~$r$). However, desirable as it is, consistency is only an asymptotic property, and offers little finite-sample guarantees:  both the  ``eigengap strategy''  and the \citet{HallinLiska07} one, for finite $n$ and depending on (b) above, may fail to separate unbounded and bounded eigenvalues: essentially, they will separate the ``rapidly diverging'' ones (e.g., linearly divergent with ``large'' slopes) from the ``slowly diverging''  (linearly divergent with ``small'' slopes) ones---the latter being  wrongly considered as bounded, i.e., idiosyncratic. Despite of consistency, this often leads to   finite-$n$ underestimation of the actual value of $q$ and undetected factors that are absorbed into the idiosyncratic components. 

A slowly diverging eigenvalue does not imply that the corresponding undetected factor or shock is negligible, though, and this is why, in Section~\ref{sec:idio}, we recommend not to throw  idiosyncratic components away, e.g. in forecasting problems.

%

\color{black}

\section{Undetected 
 factors, idiosyncratic components, orthogonality, and forecasting:  {\it ``don't throw out the baby with the bathwater!''}}\label{sec:idio}


\citet{Onatski12}'s justification for considering rate-weak factors is more subtle, though. His claim  is that weak-factor asymptotics provide a better approximation in a finite-$(n,T)$ situation where the smallest linearly divergent eigenvalues do not separate well from the bulk of bounded eigenvalues. 
He does not require, thus, the presence of a ``genuinely rate-weak'' factor,  but uses rate-weak asymptotics as a tool for improved  detection and estimation of   poorly pervasive, hence hardly detectable, strong factors.  


One may wonder, however, whether this is  worth the effort. Identifying   hardly detectable factors and incorporating them in the common space, indeed, has  limited practical impact:  undetected  factors remain undetected because their empirical finite-$(n,T)$ cross-correlations are  ``small.''  The benefits of detecting such  factors and incorporating them into the common component is that their  cross-correlations, there, would be fully exploited---which they are not in a componentwise   (no cross-correlation  exploited) or sparse  (only a few cross-correlations exploited) analysis of the idiosyncratic.  Since  these   cross-correlations are  too small to be detected, these potential  benefits are small, too, and failing to detect these factors is basically harmless. The problem of undetected factors, thus,  is  perhaps not as crucial,   in practice, as it seems---provided, however, that factor models are not considered as a dimension-reduction technique throwing away  idiosyncratic components---an attitude that is  adopted by several authors.


Whether static or dynamic, undetected factors, indeed, are not lost, but wrongly labeled as idiosyncratic. In a dimension-reduction perspective, the common component $\chi_{it}$ (be it $\chi_{it}^{\scriptscriptstyle{\text{\rm stat}}}$ or $\chi_{it}^{\scriptscriptstyle{\text{\rm dyn}}}$) is considered an   approximation of the high-dimensional observation $X_{it}$  which, being reduced rank,  is more tractable than~$X_{it}$ itself. The idiosyncratic $\xi_{it}$ (either $\xi_{it}^{\scriptscriptstyle{\text{\rm stat}}}$ or $\xi_{it}^{\scriptscriptstyle{\text{\rm dyn}}}$) then is discarded as if it were a negligible {\it error term}---a somewhat regrettable ({\it errors}, in principle, are memoryless, hence cannot be predicted) terminology used by some authors  and originating in the exact factor model literature where, by definition, no idiosyncratic lagged cross-correlation is present.  Depending on the objective of the analysis or the nature of the dataset, this may  or may not  be legitimate.   In an analysis aiming at  constructing a few indices that summarize the ``global state'' of the economy and its  impact   on individual cross-sectional items, for instance, extracting the factors and projecting observed $X_{it}$'s on the factor space (yielding the common component $\chi_{it}$)  is the obvious way to go: a dimension-reduction attitude, thus, in which idiosyncratic components safely can be ignored. The same attitude is   common practice, e.g., in macroeconometric forecasting, where it seems that  the idiosyncratic 
 safely can be neglected so that fore\-casting~$X_{i,t+h}$ boils down to forecasting the value at time $t+h$ of the common component, be it static or dynamic 
  \citep{boivin2006more,luciani2014forecasting}.



Whether static or dynamic, idiosyncratic components 
 need not be small, though, and may be   strongly autocorrelated, hence enjoying high predictive value for $X_{i, t+h}$.  
Discarding them, in a prediction context, is potentially quite damaging. This is often the case in financial applications dealing with panels of volatility measures \citep{herskovic2016common,barigozzi2017generalized,barigozzi2020generalized}.  
Rather than a dimension-reduction technique, factor models then should be considered a {\it ``divide and conquer''} procedure where the common and the idiosyncratic are analyzed via distinct appropriate methods---reduced-rank-process techniques for the common, taking advantage of  strong and pervasive cross-correlations, and componentwise or sparse techniques for the idiosyncratic, where little or no advantage can be gained from the negligible potential cross-correlations. 

In forecasting problems,
 depending on the type  (static or dynamic) of factor model adopted, this ``divide and conquer'' approach yields $h$-step ahead forecasts~${\chi}_{i,t+h\vert t}^{\scriptscriptstyle\text{\rm stat}} $ (based on the   past~$\{\chi_{js}^{\scriptscriptstyle{\text{\rm stat}}}\vert\,  j\in{\mathbb N}, s\leq~\!t\}$\linebreak of all common components)  of~$\chi_{i, t+h}^{\scriptscriptstyle\text{\rm stat}}$ and~${\xi}_{i,t+h\vert t}^{\scriptscriptstyle\text{\rm stat}}$ (based, e.g.,  on the past $\{\xi_{is}^{\scriptscriptstyle\text{\rm stat}}\vert\,   s\leq t\}$ of the $i$th idio\-syncratic component only) 
 of  $\xi_{i, t+h}^{\scriptscriptstyle\text{\rm stat}}$, or fore\-casts~${\chi}_{i,t+h\vert t}^{\scriptscriptstyle\text{\rm dyn}} $ (based on~$\{\chi_{js}^{\scriptscriptstyle{\text{\rm dyn}}}\vert\,  j\in{\mathbb N}, s\leq t\}$)  of~$\chi_{i, t+h}^{\scriptscriptstyle\text{\rm dyn}}$ and~${\xi}_{i,t+h\vert t}^{\scriptscriptstyle\text{\rm dyn}}$  (based   on 
$\{\xi_{is}^{\scriptscriptstyle\text{\rm sdyn}}\vert\,   s\leq t\}$) of $\xi_{i, t+h}^{\scriptscriptstyle\text{\rm dyn}}$. In these examples, a componentwise univariate forecasting strategy is considered for  idiosyncratic components. More efficient idiosyncratic prediction strategies   could be adopted, where several past idiosyncratic values are taken into account (as in  sparse VAR prediction methods); in theory, one even could use all past idiosyncratic values, which in practice is rapidly infeasible, though. Mutatis mutandis,   the results below would remain unchanged. For  simplicity of exposition and notation, we therefore restrict our predictors of ${\xi}_{i,t+h}^{\scriptscriptstyle\text{\rm stat}}$ and ${\xi}_{i,t+h}^{\scriptscriptstyle\text{\rm dyn}}$ to the componentwise forecasts $\xi_{i, t+h}^{\scriptscriptstyle\text{\rm stat}}$ and $\xi_{i, t+h}^{\scriptscriptstyle\text{\rm dyn}}$ just described.

After being obtained separately, however, these forecasts  
 somehow should  be brought back together to produce a forecast of $X_{i, t+h}$. To this end, one naturally 
  could consider their   sums 
  \begin{equation}\label{sumstat}
  {X}_{i, t+h\vert t}^{\scriptscriptstyle\text{\rm stat}}\coloneqq {\chi}_{i,t+h\vert t}^{\scriptscriptstyle\text{\rm stat}}  + {\xi}_{i,t+h\vert t}^{\scriptscriptstyle\text{\rm stat}}
  \end{equation} 
  and
   \begin{equation}\label{sumdyn}{X}_{i, t+h\vert t}^{\scriptscriptstyle\text{\rm dyn}}\coloneqq {\chi}_{i,t+h\vert t}^{\scriptscriptstyle\text{\rm dyn}}  + {\xi}_{i,t+h\vert t}^{\scriptscriptstyle\text{\rm dyn}}.
  \end{equation} 
      as  pre\-dictors for~$X_{i,t+h}$. 
   This   is what   \citet{fan2023bridging} are doing in the context of  {static approximate}  factor models and  \citet{barigozzi2024fnets} in  the GDFM context, with sparse VAR empirical fore\-casts~$\widehat{\xi}_{i,t+h\vert t}^{\scriptscriptstyle\text{\rm stat}}$ and~$\widehat{\xi}_{i,t+h\vert t}^{\scriptscriptstyle\text{\rm dyn}}$ of ${\xi}_{i,t+h}^{\scriptscriptstyle\text{\rm stat}}$ and ${\xi}_{i,t+h}^{\scriptscriptstyle\text{\rm dyn}}$, respectively.

    The following result shows that while 
   ${X}_{i, t+h\vert t}^{\scriptscriptstyle\text{\rm stat}}$ 
   is not necessarily the best linear predictor of~$X_{i,t+h}$ based on the present and past of 
   $\{(\chi_{jt}^{\scriptscriptstyle{\text{\rm stat}}}, \xi_{it}^{\scriptscriptstyle{\text{\rm stat}}})\}$,   ${X}_{i, t+h\vert t}^{\scriptscriptstyle\text{\rm dyn}}$ is the best linear predictor of~$X_{i,t+h}$ based on the present and past of 
   $\{(\chi_{jt}^{\scriptscriptstyle{\text{\rm dyn}}}, \xi_{it}^{\scriptscriptstyle{\text{\rm dyn}}})\}$ and a.s.\ outperforms ${X}_{i, t+h\vert t}^{\scriptscriptstyle\text{\rm stat}}$. {\it Best}, here and in the sequel, is to be understood in the classical sense of minimal expected squared prediction error.

Denote by ${\mathcal H}^{\bf X}_{t{\rceil}}$,  
${\mathcal H}^{{\boldsymbol\chi}}_{t{\rceil}\scriptscriptstyle{\text{\rm stat}}}$, 
${\mathcal H}^{{\boldsymbol\chi}}_{t{\rceil}\scriptscriptstyle{\text{\rm dyn}}}$,   
 ${\mathcal H}^{{\xi}_i}_{t{\rceil}\scriptscriptstyle{\text{\rm stat}}}$,  
  ${\mathcal H}^{{\xi}_i}_{t{\rceil}\scriptscriptstyle{\text{\rm dyn}}}$, 
  ${\mathcal H}_{t{\rceil}\scriptscriptstyle{\text{\rm stat}}}^{{\boldsymbol\chi}\, {\xi_i}}$, and  
${\mathcal H}_{t{\rceil}\scriptscriptstyle{\text{\rm dyn}}}^{{\boldsymbol\chi}\, {\xi_i}}$ 
the Hilbert spaces spanned\linebreak 
 by~$\{X_{is}\vert\, i\in{\mathbb N}, s\leq t\}$, 
$\{\chi_{js}^{\scriptscriptstyle{\text{\rm stat}}}\vert\,  j\in{\mathbb N}, s\leq t\}$, 
$\{\chi_{js}^{\scriptscriptstyle\text{\rm dyn}}\vert\,  j\in{\mathbb N}, s\leq t\}$, 
$\{\xi_{is}^{\scriptscriptstyle\text{\rm stat}}\vert\,   s\leq t\}$, 
$\{\xi_{is}^{\scriptscriptstyle\text{\rm dyn}}\vert\,   s\leq t\}$, 
$\{(\chi_{js}^{\scriptscriptstyle{\text{\rm stat}}}, \xi_{is}^{\scriptscriptstyle{\text{\rm stat}}})\vert\, j\in{\mathbb N}, s\leq t\}$, and 
$\{(\chi_{js}^{\scriptscriptstyle{\text{\rm dyn}}}, \xi_{is}^{\scriptscriptstyle{\text{\rm dyn}}})\vert\, j\in{\mathbb N}, s\leq t\}$, 
respectively. If  componentwise forecasts are adopted for  idiosyncratic components, the best linear $h$-step ahead predictors of $X_{t+h\vert t}$ based on the static approximate factor model  and the GDFM decompositions, are the projections 
$$
{X}_{i, t+h\vert t}^{{\scriptscriptstyle\text{\rm stat}}*}\coloneqq \text{\rm Proj} \left({X}_{i, t+h} \vert\, {\mathcal H}_{t{\rceil}\scriptscriptstyle{\text{\rm stat}}}^{{\boldsymbol\chi}\, {\xi_i}}
\right)\quad\text{and}\quad {X}_{i, t+h\vert t}^{{\scriptscriptstyle\text{\rm dyn}}*}\coloneqq \text{\rm Proj} \left({X}_{i, t+h} \vert\, {\mathcal H}_{t{\rceil}\scriptscriptstyle{\text{\rm dyn}}}^{{\boldsymbol\chi}\, {\xi_i}}
\right),
$$
 respectively. For the common and  idiosyncratic components, these best linear predictors  are 
 $$
 {\chi}_{i,t+h\vert t}^{\scriptscriptstyle\text{\rm stat}} \coloneqq \text{\rm Proj} \left({\chi}_{i, t+h}^{\scriptscriptstyle{\text{\rm stat}}}\ \vert\, {\mathcal H}_{t{\rceil}\scriptscriptstyle{\text{\rm stat}}}^{{\boldsymbol\chi}}
\right)
 ,\qquad\ \ \ \  {\xi}_{i,t+h\vert t}^{\scriptscriptstyle\text{\rm stat}} \coloneqq   \text{\rm Proj} \left({\xi}_{i, t+h}^{\scriptscriptstyle{\text{\rm stat}}}\ \vert\, {\mathcal H}_{t{\rceil}\scriptscriptstyle{\text{\rm stat}}}^{{\xi}_i}
\right) ,
$$
 $$
{\chi}_{i,t+h\vert t}^{\scriptscriptstyle\text{\rm dyn}} \coloneqq
 \text{\rm Proj} 
 \left(
 {\chi}_{i, t+h}^{\scriptscriptstyle{\text{\rm dyn}}} \vert\,  
{\mathcal H}_{t{\rceil}
 \scriptscriptstyle{\text{\rm dyn}}}^{{\boldsymbol\chi}}
\right),
 \quad\text{and}\quad 
  {\xi}_{i,t+h\vert t}^{\scriptscriptstyle\text{\rm dyn}} \coloneqq   \text{\rm Proj} \left(
  {\xi}_{i, t+h}^{\scriptscriptstyle{\text{\rm dyn}}}\ \vert\, {\mathcal H}_{t{\rceil}\scriptscriptstyle{\text{\rm dyn}}}^{{\xi}_i}
\right)
 .$$

Now, 
 since ${\chi}_{i, t+h} ^{\scriptscriptstyle\text{\rm dyn}} $ is orthogonal to all $\xi_{is}^{\scriptscriptstyle\text{\rm dyn}}$'s, hence to ${\mathcal H}^{{\xi}_i}_{t{\rceil}\scriptscriptstyle{\text{\rm dyn}}}$,  and since  $\xi_{it}^{\scriptscriptstyle\text{\rm dyn}}$ is orthogonal to all ${\chi}_{js} ^{\scriptscriptstyle\text{\rm dyn}} $'s, hence to ${\mathcal H}^{{\boldsymbol\chi}}_{t{\rceil}\scriptscriptstyle{\text{\rm dyn}}}$, 
\begin{eqnarray*}
{X}_{i, t+h\vert t}^{{\scriptscriptstyle\text{\rm dyn}}*} &=&
\text{\rm Proj} \left({\chi}_{i, t+h} ^{\scriptscriptstyle\text{\rm dyn}}  \vert\, {\mathcal H}_{t{\rceil}\scriptscriptstyle{\text{\rm dyn}}}^{{\boldsymbol\chi}\, {\xi_i}}
\right) + \text{\rm Proj} \left({\xi}_{i, t+h} ^{\scriptscriptstyle\text{\rm dyn}}  \vert\, {\mathcal H}_{t{\rceil}\scriptscriptstyle{\text{\rm dyn}}}^{{\boldsymbol\chi}\, {\xi_i}}
\right) \\ 
&=& 
\text{\rm Proj} \left({\chi}_{i, t+h} ^{\scriptscriptstyle\text{\rm dyn}}  \vert\, {\mathcal H}_{t{\rceil}\scriptscriptstyle{\text{\rm dyn}}}^{{\boldsymbol\chi}}
\right) +  \text{\rm Proj} \left({\xi}_{i, t+h} ^{\scriptscriptstyle\text{\rm dyn}}  \vert\, {\mathcal H}_{t{\rceil}\scriptscriptstyle{\text{\rm dyn}}}^{{\xi}_i}\right)
= {\chi}_{i,t+h\vert t}^{\scriptscriptstyle\text{\rm dyn}}  + {\xi}_{i,t+h\vert t}^{\scriptscriptstyle\text{\rm dyn}} = {X}_{i, t+h\vert t}^{{\scriptscriptstyle\text{\rm dyn}}} .
\end{eqnarray*}
In the GDFM approach, adding (as in~\eqref{sumdyn}) the optimal predictor ${\chi}_{i,t+h\vert t}^{\scriptscriptstyle\text{\rm dyn}} $ of ${\chi}_{i, t+h}^{\scriptscriptstyle\text{\rm dyn}} $ and the optimal marginal predictor ${\xi}_{i,t+h\vert t}^{\scriptscriptstyle\text{\rm dyn}}$ of ${\xi}_{i, t+h} ^{\scriptscriptstyle\text{\rm dyn}} $, thus, yields the best linear predictor ${X}_{i, t+h\vert t}^{{\scriptscriptstyle\text{\rm dyn}}*} $: this additive combination of common and idiosyncratic forecasts, therefore, is  fully justified. This is not the case in a static approximate approach: ${\chi}_{i, t+h} ^{\scriptscriptstyle\text{\rm stat}} $, indeed, is  orthogonal to  $\xi_{i, t+h}^{\scriptscriptstyle\text{\rm stat}}$ only, hence fails, in general,  to be ortho\-gonal to ${\mathcal H}^{{\xi}_i}_{t{\rceil}\scriptscriptstyle{\text{\rm stat}}}$. Similarly, $\xi_{it}^{\scriptscriptstyle\text{\rm stat}}$ is orthogonal to   ${\chi}_{jt} ^{\scriptscriptstyle\text{\rm stat}} $ for all $j\in{\mathbb N}$ but not to ${\mathcal H}^{{\boldsymbol\chi}}_{t{\rceil}\scriptscriptstyle{\text{\rm stat}}}$. As a consequence, adding ${\chi}_{i,t+h\vert t}^{\scriptscriptstyle\text{\rm stat}} $ and  ${\xi}_{i,t+h\vert t}^{\scriptscriptstyle\text{\rm stat}}$, as a rule, does not yield the best linear predictor ${X}_{i, t+h\vert t}^{{\scriptscriptstyle\text{\rm stat}}*} $.

Moreover, 
 ${X}_{i, t+h\vert t}^{\scriptscriptstyle\text{\rm dyn}} = {X}_{i, t+h\vert t}^{{\scriptscriptstyle\text{\rm dyn}}*} $ outperforms the (hardly implementable)~${X}_{i, t+h\vert t}^{{\scriptscriptstyle\text{\rm stat}}*} $ which, in turn, outperforms the suboptimal ${X}_{i, t+h\vert t}^{{\scriptscriptstyle\text{\rm stat}}}= {\chi}_{i,t+h\vert t}^{\scriptscriptstyle\text{\rm stat}} +{\xi}_{i,t+h\vert t}^{\scriptscriptstyle\text{\rm stat}}$. Indeed,  
${\mathcal H}_{t{\rceil}\scriptscriptstyle{\text{\rm dyn}}}^{{\boldsymbol\chi}\, {\xi_i}}$, which contains   ${\chi}_{js}^{\scriptscriptstyle\text{\rm weak}} $ for all~$j\in{\mathbb N}$ and~$s\leq t$, includes ${\mathcal H}_{t{\rceil}\scriptscriptstyle{\text{\rm stat}}}^{{\boldsymbol\chi}\, {\xi_i}}$, which only contains   ${\chi}_{is}^{\scriptscriptstyle\text{\rm weak}} $, $s\leq t$. Hence, ${\mathcal H}_{t{\rceil}\scriptscriptstyle{\text{\rm dyn}}}^{{\boldsymbol\chi}\, {\xi_i}} \supseteq {\mathcal H}_{t{\rceil}\scriptscriptstyle{\text{\rm stat}}}^{{\boldsymbol\chi}\, {\xi_i}}$ and the squared prediction error $\left[ X_{i, t+h} - {X}_{i, t+h\vert t}^{\scriptscriptstyle\text{\rm dyn}}\right]^2$ is a.s.\  less than  or equal to the (generally unimple\-mentable)~$\left[ X_{i, t+h} - \wh{X}_{i, t+h\vert t}^{{\scriptscriptstyle\text{\rm stat}}*}\right]^2\!$, which in turn is a.s.\  less than  or equal to $\left[ X_{i, t+h}  - {X}_{i, t+h\vert t}^{{\scriptscriptstyle\text{\rm stat}}}\right]^2$.\smallskip

We have thus proved the following result.
\begin{prop}\label{Th2} Let $\bf X$ be second-order stationary and satisfy both \eqref{7} and \eqref{9}. Then, 
 \begin{enumerate}
 \item[(i)] ${X}_{i, t+h\vert t}^{{\scriptscriptstyle\text{\rm stat}}}$ defined in~\eqref{sumstat}  is not, in general, the best $h$-step ahead  linear predictor of $X_{i, t+h}$ based on the present and past of $\{(\chi_{jt}^{\scriptscriptstyle{\text{\rm stat}}}, \xi_{it}^{\scriptscriptstyle{\text{\rm stat}}})\vert\, j\in{\mathbb N}\}$;
 \item[(ii)] ${X}_{i, t+h\vert t}^{{\scriptscriptstyle\text{\rm dyn}}}$ defined in~\eqref{sumdyn}  is   the best $h$-step ahead  linear predictor of $X_{i, t+h}$ based on the present and past of  $\{(\chi_{jt}^{\scriptscriptstyle{\text{\rm dyn}}}, \xi_{it}^{\scriptscriptstyle{\text{\rm dyn}}})\vert\, j\in{\mathbb N}\}$;
 \item[(iii)] ${X}_{i, t+h\vert t}^{{\scriptscriptstyle\text{\rm dyn}}}$ outperforms the best  $h$-step ahead  linear predictor of $X_{i, t+h}$ based on the present and past of $\{(\chi_{jt}^{\scriptscriptstyle{\text{\rm stat}}}, \xi_{it}^{\scriptscriptstyle{\text{\rm stat}}})\vert\, j\in{\mathbb N}\}$, hence also outperforms ${X}_{i, t+h\vert t}^{{\scriptscriptstyle\text{\rm stat}}}$.
 \end{enumerate}
 \end{prop}

%


Similar statements could be made for predictors ${X}_{i, t+h\vert t}^{{\scriptscriptstyle\text{\rm dyn}}}$ and ${X}_{i, t+h\vert t}^{{\scriptscriptstyle\text{\rm stat}}}$ based on  alternative idiosyncratic prediction strategies (such as sparse VAR prediction): irrespective of that  strategy,  the key reason explaining the superiority of ${X}_{i, t+h\vert t}^{{\scriptscriptstyle\text{\rm dyn}}}$ over ${X}_{i, t+h\vert t}^{{\scriptscriptstyle\text{\rm stat}}}$ is the mutual orthogonality {\it at all leads and lags} of the  dynamic common and idiosyncratic components.  Yet another reason for adopting the GDFM rather than the Chamberlain and Rothschild static approximate factor   approach!

\color{black}

\section{Conclusions}\label{concSec}

This paper is dealing with several fundamental issues in the theory and practice of factor models.  Its main conclusions   are as follows.
\begin{enumerate}
\item[(i)]  Since (as shown by \cite{GERSTH23} and  \citet{Gersingetal23})  it is nesting the classical  and widespread static   model of  \cite{Chamberlain83} and \cite{ChamberlainRothschild83}, the General Dynamic Factor Model of \cite{FHLR00}  is uniformly preferable; the benefit is the incorporation into the common space of the GRD-weak factors which consist of  statically non-pervasive 
 leads and lags   of statically strong factors and, in the static approach, are treated as idiosyncratic; 

\item[(ii)] under the (natural) assumption of cross-sectional exchangeability, {rate-weak and rate-superstrong  factors}, whether static or dynamic, {``are nowhere''};

\item[(iii)] {\it   Gersing-Rust-Deistler-weak factors ``are everywhere,''} but  in the GDFM approach are incorporated into the space of  (dynamically) common components, where they can be efficiently exploited---which explains the empirical finding that the {dynamic} approach outperforms the {static} one even when the assumptions of the static factor model are satisfied;  

\item[(iv)] in forecasting problems, idiosyncratic  components (which are not  {\it error terms}) should not be\linebreak  systematically  discarded, as they may be large and have high predictive power; unless ortho\-gonality at all leads and lags  is imposed between the statically common and idiosyncratic components (which is a very strong assumption),  efficiently combining  their  predictors into   predictors for $\bf X$, however, is not obvious while, in sharp contrast, summing  dynamically common   and dynamically idiosyncratic predictors (which are mutually ortho\-gonal at all leads and lags) yields optimal linear predictors for $\bf X$. 
\end{enumerate}

These conclusions about the advantages of the GDFM over the approximate static model of  Chamberlain and Rothschild 
are likely to extend to spatio-temporal data (for which a GDFM representation result exists, see \cite{BLVL23}), to locally stationary and  piecewise stationary high-dimensional time series (for which a GDFM approach is available, see \citet{barigozzi2021time} and, for the piecewise stationary case,   \citet{cho2024high}, without representation result, though), to functional ones (see \cite{HTava23}, where a static representation result is established), and to high-dimensional  tensor-valued time series (for which a dynamic approach similar to the GDFM still needs to be developed).

%

\bibliographystyle{chicago}
\bibliography{Factors.bib}

\bigskip\bigskip

\noindent{\Large\sc \bf Appendix:  Proof of Proposition~1}\bigskip



The classical  Weyl inequalities straightforwardly entail that (B) implies (A). 
The proof of Proposition~1, thus, essentially consists of establishing   that (A)---that is,  the assumption \eqref{9} on the eigen\-values~$\lambda^{(n)}_{X;j}$ 
 of the covariance matrices ${\boldsymbol{\Gamma}}^{(n)}_{X}\coloneqq {\rm E}[{\bf X}^{(n)}_t {\bf X}^{(n)\prime}_t]$ of~${\bf X}^{(n)}_t\coloneqq (X_{1t},\ldots,X_{nt})\pr$ for~$n\in{\mathbb N}$---implies that the conditions (a)--(f) on the static approximate factor model decomposition~\eqref{2} of Chamberlain and Rothschild (along with (d$_2$)) are satisfied by the canonical decomposition \eqref{8}. Essentially, this consists in showing that, under~(A),  
 the covariance matrices ${\boldsymbol\Gamma}\n_{\chi}\coloneqq  {\rm E}[{\boldsymbol\chi}^{(n)}_t {\boldsymbol\chi}^{(n)\prime}_t]$  and ${\boldsymbol\Gamma}\n_{\xi}\coloneqq  {\rm E}[{\boldsymbol\xi}^{(n)}_t {\boldsymbol\xi}^{(n)\prime}_t]$ of the statically common and idiosyncratic components~${\boldsymbol\chi}\n_t\coloneqq (\chi_{1t}^{\scriptscriptstyle\text{\rm stat}},\ldots, \chi_{nt}^{\scriptscriptstyle\text{\rm stat}})\pr$ and~${\boldsymbol\xi}\n_t \coloneqq (\xi_{1t}^{\scriptscriptstyle\text{\rm stat}},\ldots, \xi_{nt}^{\scriptscriptstyle\text{\rm stat}})\pr$ defined in  \eqref{8} and their eigenvalues are such that \vspace{-1mm}
\begin{enumerate}
\item[(i)] $\chi_{it}^{\scriptscriptstyle\text{\rm stat}}$  is of the form ${\bf B}_i{\bf f}_t$ for some  $1\times r$ vector ${\bf B}_i$ and some $r\times 1$  random vector ${\bf f}_t$ satisfying assumption (c) in Section 2.1; \vspace{-1mm}
\item[(ii)] the $r$ non-zero eigenvalues $\lambda\n_{\chi ;1}, \ldots , \lambda\n_{\chi ;r}$ of ${\boldsymbol\Gamma}\n_{\chi}$ tend to infinity as $n\to\infty$ (it readily follows from~(i) that~$\lambda\n_{\chi;j}=0$ for all $r+1\leq  j\leq n$);\vspace{-1mm}
\item[(iii)] the first (largest) eigenvalue $\lambda\n_{\xi;1}$ of ${\boldsymbol\Gamma}\n_{\xi}$ is bounded as $n\to\infty$.
\end{enumerate}
This takes care, for \eqref{8}, of conditions (b)--(e), while conditions (a) and (f) are automatically satisfied in view of the definition of $\chi_{it}^{\scriptscriptstyle\text{\rm stat}} $ and $\xi_{it}^{\scriptscriptstyle\text{\rm stat}}$ in \eqref{8}. The uniqueness of the decomposition satisfying (a)--(f) then entails that \eqref{8} and \eqref{2} coincide.

Below,  {to keep the notation simple,} we denote by $\mathcal H$, $\mathcal{H}_{\scriptscriptstyle\text{\rm com}}$, and $\mathcal{H}_{\scriptscriptstyle\text{\rm idio}}$ the Hilbert spaces $\mathcal{H}^{{\bf X}_t}$,~$\mathcal{H}_{\scriptscriptstyle\text{\rm stat com}}^{{\bf X}_t}$, and $\mathcal{H}_{\scriptscriptstyle\text{\rm stat idio}}^{{\bf X}_t}$ of Section~\ref{Sec4}.  The unit vector $z/\Vert z\Vert$ corresponding to $0\neq z\in\mathcal H$ (the standardized version~$z/\sigma_z$ of $z$ with variance~$\sigma_z^2$) is denoted by $z_*$. 
Denoting by $w\n_1,\ldots, w\n_n$ the principal components of ${\boldsymbol{\Gamma}}^{(n)}_{X}$, write ${\mathcal H}\n_{r\rceil}$ for the (closed)~$r$-dimensional subspace of~$\mathcal H$ spanned by~$w\n_1,\ldots, w\n_r$  and  ${\mathcal H}\n_{\lceil r+1}$ for the (closed) subspace of~$\mathcal H$ spanned by~$w\n_{r+1}, w\n_{r+2},\,\ldots $ 

Recall that an infinite-dimensional Hilbert space ${\mathcal H}$ admits two distinct topologies, the {\it weak} topology  and the {\it strong} one. In the strong topology,  convergence is associated with the norm:~$ z\n$ converges strongly to $ z$ (notation: $ z\n\to z$) if $\Vert
 z\n -  z\Vert \to 0$ as $n\to\infty$---strong convergence, in our case, thus, is convergence in quadratic mean (q.m.~convergence). Weak convergence of $ z\n$ to $ z$ (nota\-tion:~$ z\n\rightharpoonup z$) is characterized by the fact that~$\langle  z\n , z\rangle \to \langle  z , z\rangle$ (in our case, Cov$( z\n , z) \to$ Cov$( z , z) $)  for any~$z\in{\mathcal H}$ as $n\to\infty$ for any~$z\in{\mathcal H}$ as~$n\to\infty$. For finite-dimensional Hilbert spaces, the weak and strong topologies coincide. An important (Bolzano-Weierstrass-type)  property is that any bounded  sequence~$ z\n$ (bounded in the norm---in our case, Var($ z\n$) bounded) admits a subsequence\footnote{Any subsequence of a sequence of elements of $\mathcal H$, of course, still is a sequence of elements of $\mathcal H$. } converging weakly to some $ z$---see, e.g., section~D.4 on   page~639 of~\cite{Evans10} or Theorem~3.18 in \cite{Brezis11}. Finally, recall that orthogonal projections, in $\mathcal H$, are linear continuous operators. 

The proof is organised in four steps.

(1) We first characterize ${\mathcal H}_{\scriptscriptstyle{\text{\rm   com}}}$ in terms of the limits of sequences of standardized elements of ${\mathcal H}\n_{r\rceil}$.
By definition, the elements~$\zeta\neq 0$ (with variance $\sigma^2_\zeta $) of ${\mathcal H}_{\scriptscriptstyle{\text{\rm   com}}}$ are such that $\zeta_*$ is the strong limit  (the q.m.~limit), as~$n\to\infty$, of sequences  $(w\n_*)_{n\in{\mathbb{N}}}$ where  $w\n\coloneqq \sum_{i=1}^n c\n_iX_{it}$ with   $\sum_{i=1}^n (c\n_i)^2 = 1$ has variance  $(\sigma\n_w)^2 \to \infty$ as $n\to\infty$. Note that $\sum_{i=1}^n (c\n_i)^2 = 1$, in this definition, can be replaced with~$\sum_{i=1}^n (c\n_i)^2 = (c\n)^2>0$ provided that $(c\n)^2$ 
remains bounded away from zero and infinity. Denote by ${\mathcal W}_*\coloneqq \big\{ (w\n_*)_{n\in\mathbb{N}}\big\}$ the collection of such  sequences: $0\neq \zeta\in {\mathcal H}_{\scriptscriptstyle{\text{\rm   com}}}$, thus, iff $\zeta_*$ is the strong limit of a sequence $(w\n_*)_{n\in{\mathbb{N}}}\in{\mathcal W}_*$. Clearly, for~$j=1,\ldots, r$, the sequen\-ces~$(w\n_{j*})_{n\in{\mathbb N}}$ of standardized principal components  of ${\boldsymbol{\Gamma}}^{(n)}_{X}$, hence the sequences of  standardized elements of  ${\mathcal H}\n_{r\rceil}$,  belong to ${\mathcal W}_*$. Denoting by ${\mathcal W}_{r\rceil *}\subseteq{\mathcal W}_*$ the collection of such sequences, let us show that $0\neq \zeta\in{\mathcal H}_{\scriptscriptstyle{\text{\rm   com}}}$ iff $\zeta_*$ is the strong limit   of a sequence $(w\n_*)_{n\in{\mathbb{N}}}\in{\mathcal W}_{r\rceil *}$, that is, \vspace{-1.5mm}
$${\mathcal H}_{\scriptscriptstyle{\text{\rm   com}}}
=
\left\{z= \sigma \lim_{n\to\infty} w\n_*
{\big\vert}\, (w\n_*)_{n\in{\mathbb{N}}}\in{\mathcal W}_{r\rceil *}
,\, \sigma\in{\mathbb R}_+\right\}.
\vspace{-1.5mm}$$
(note that $\lim_{n\to\infty} w\n_*$ can be replaced with $\lim_{k\to\infty} w^{n_k}_*$ where $(w^{n_k}_*)_{k\in{\mathbb N}}$ is a subsequence of $(w^{n}_*)_{n\in{\mathbb N}}$).

To show this, consider $(w\n)_{n\in\mathbb{N}}$  such that $(w\n_*)_{n\in\mathbb{N}}\in {\mathcal W}_*$. Each $w_n$ canonically decomposes into a sum 
 $w\n = w\n_{r\rceil} +~\!w\n_{\lceil r+1}$ with~$w\n_{r\rceil}$ and~$w\n_{r\rceil *} \in {\mathcal H}\n_{r\rceil}$ orthogonal to $w\n_{\lceil r+1 }$ and $w\n_{\lceil r+1 *} \in{\mathcal H}\n_{\lceil r+1}$; the variances  $(\sigma\n_{w_{r\rceil}})^2$ tend to infinity while the variances  $(\sigma\n_{w_{\lceil r+1}} )^2$ remain bounded as~$n\to\infty$. Hence, if~$\zeta_*$ is the strong~limit of a se\-quence~$(w\n_*)_{n\in\mathbb{N}}$ of ${\mathcal W}_*$,   it is also the strong~limit of the sequence $(w\n_{r\rceil *})_{n\in\mathbb{N}}$ of   
 ${\mathcal W}_{r\rceil *}$.  
 
 (2) Next,  we show that a subsequence of  the $r$-tuple $(w\n_{1*},\ldots,w\n_{r*})_{n\in{\mathbb N}}$ converges weakly to an orthogonal system $(w_{1*},\ldots,w_{r*})$. Since for each $n$ the $r$-tuple $w\n_{1*},\ldots,w\n_{r*}$ is an orthonormal basis of~${\mathcal H}\n_{r\rceil}$, any~$(w\n_*)_{n\in{\mathbb N}}$ in ${\mathcal W}_{r\rceil *}$ is such that~$w\n_*$ is a linear combination of $w\n_{1*},\ldots,w\n_{r*}$. As \linebreak an $r$-tuple of  bounded sequence,$(w\n_{1*},\ldots,w\n_{r*})_{n\in{\mathbb N}}$ contains a  subsequence $(w^{(n_k)}_{1*},\ldots,w^{(n_k)}_{r*})_{k\in{\mathbb N}}$ converging weakly to some $r$-tuple $(w_{1*},\ldots,w_{r*})$. Weak convergence implies the convergence of covariances, and hence preserves orthogonality: indeed,
 {\begin{align*}  \lim_{k\to\infty} \text{Cov}(w_{j_1*},w_{j_2*})&=\lim_{k\to\infty} \text{Cov}(w^{(n_k)}_{j_1*},w_{j_2*})=\lim_{k\to\infty} \text{Cov}(w^{(n_k)}_{j_1*},w^{(n_k)}_{j_2*} +(w_{j_2*}-w^{(n_k)}_{j_2*})) \\ 
  &= 0 + \lim_{k\to\infty} \text{Cov}(w^{(n_k)}_{j_1*}, o_{\text{q.m.}}(1)) = 0.
  \end{align*}
  The $r$-tuple $(w_{1*},\ldots,w_{r*})$, thus, is an orthogonal  basis of the~$r$-dimensional space ${\mathcal H}_{r\rceil}$ it is spanning. 
  
  (3) Let us  conclude that  ${\mathcal H}_{\scriptscriptstyle{\text{\rm   com}}}$ coincides with ${\mathcal H}_{r\rceil}$, hence has finite dimension $r$.  Denote by Proj$_{{\mathcal H}_{r\rceil}}$ the orthogonal projection operator from $\mathcal H$ to ${\mathcal H}_{r\rceil}$. As a projection, the mapping Proj$_{{\mathcal H}_{r\rceil}}$ is linear and continuous. Hence,
   \begin{equation*}
  w^{(n_k)}_{j*}\rightharpoonup w_{j*}\text{  implies }\text{Proj}_{{\mathcal H}_{r\rceil}}  w^{(n_k)}_{j*}\rightharpoonup 
  \text{Proj}_{{\mathcal H}_{r\rceil}}w_{j*} =w_{j*},\quad j=1,\ldots,r\quad  \text{as $k\to\infty$}  
  \end{equation*}
  where the latter convergence takes place in a finite-dimensional space, so that it also holds in the strong topology: $\text{Proj}_{{\mathcal H}_{r\rceil}}  w^{(n_k)}_{j*}\to  
 w_{j*}$. Hence, the   strong limits $w_*$ of sequences $(w^{(n_k)}_*)_{k \in {\mathbb{N}}}\in{\mathcal W}_{r\rceil *}$ also  are strong limits of  the sequence of  their projections  on ${{\mathcal H}_{r\rceil}}$, and so are the   limits of strongly convergent sequences thereof. Since strong convergence implies convergence of the norms and Var($w^{(n_k)}_*$)=1 for all~$n_k$, Var($w_*$) =1 as well.  Conversely, any element in ${{\mathcal H}_{r\rceil}}$ being a linear combination of $w_{1*}, \ldots,w_{r*}$ (or the limit of a strongly convergent sequence thereof), it is  the strong limit of the sequence of the projections  
 on ${{\mathcal H}_{r\rceil}} $ of the same linear combinations of~$w_{1*}^{(n_k)}, \ldots,w_{r*}^{(n_k)}$ (or the limit of strongly convergent sequence thereof). 
 As a consequence,     $z\in{\mathcal H}_{\scriptscriptstyle{\text{\rm   com}}}$ iff $z_*$, hence $z$ itself, is an element of ${{\mathcal H}_{r\rceil}}$:  we thus have~${\mathcal H}_{\scriptscriptstyle{\text{\rm   com}}}
={{\mathcal H}_{r\rceil}}
$. 

(4) The static factor model decomposition of $X_{it}$ follows as a corollary. Denote by ${\bf f}_t=(f_{1t},\ldots ,{f}_{rt})^\prime$ (and call static factors) an arbitrary orthonormal basis of  the~$r$-dimensional space ${{\mathcal H}_{r\rceil}}$. Since~$\chi_{it}^{\scriptscriptstyle\text{\rm stat}}\in{\mathcal H}_{\scriptscriptstyle{\text{\rm   com}}}$ 
 and~${\mathcal H}_{\scriptscriptstyle{\text{\rm   com}}}
={{\mathcal H}_{r\rceil}}
 $, 
 $\chi_{it}^{\scriptscriptstyle\text{\rm stat}}= {\bf B}_i{\bf f}_t$ for ${\bf B}_i\coloneqq (B_{i1},\ldots,B_{ir})$  with~$B_{ij}\coloneqq \text{Cov}(\chi_{it}^{\scriptscriptstyle\text{\rm stat}}, f_{jt})$, $i,j \in{\mathbb N}$. This takes care of (i).

(5) To conclude, let us show  that (ii) ${\boldsymbol\Gamma}\n_{\chi}$ has exactly $r$ divergent eigenvalues while (iii)  the eigenvalues of~${\boldsymbol\Gamma}\n_{\xi}$ all are bounded. 
Let us assume   that the number of    diverging eigenvalues of ${\boldsymbol\Gamma}\n_{\chi}$ is  strictly less than $r$. Since ${\boldsymbol\Gamma}\n_X$ decomposes into ${\boldsymbol\Gamma}\n_\chi + {\boldsymbol\Gamma}\n_\xi$ and has $r$ diverging eigenvalues, the Weyl inequalities imply that the first eigenvalue $\lambda\n_{\xi; 1}$ of ${\boldsymbol\Gamma}\n_\xi$ tends to infinity as~$n\to\infty$. Then, there exists a sequence~$(w\n_\xi)_{n\in{\mathbb N}}$ with~$w\n_\xi\coloneqq \sum_{i=1}^nc\n_i\xi_{it}\in{\mathcal H}_{\scriptscriptstyle{\text{\rm   idio}}}$ and~$\sum_{i=1}^n(c\n_i)^2 =1$ such that the variance~$(\sigma\n_\xi)^2$ of $w\n_\xi$ tends to infinity as~$n\to\infty$. Since $\xi_{it}$ is the projection of $X_{it}$ on~${\mathcal H}_{\scriptscriptstyle{\text{\rm   idio}}}$,~$w\n_\xi$ also is a linear combination  $ \sum_{i=1}^nd\n_iX_{it}$ of   $X_{1t},\ldots, X_{nt}$ and (the eigenvalues of a nondegenerate projection operator are either~1 or zero, and cannot  all be zero)  $0<C^- \leq \sum_{i=1}^n(d\n_i)^2 \leq 1$. Hence, the sequence~$(w\n_{\xi *}\coloneqq w\n_{\xi}/\sigma\n_\xi)_{n\in{\mathbb N}}$, where each $w\n_{\xi *}$ has variance one, is in ${\mathcal W}_*$. This bounded sequence contains a subsequence converging weakly to some~$w_{\xi *}$; since each $w\n_{\xi *}$ is in ${\mathcal H}_{\scriptscriptstyle{\text{\rm   idio}}}$, hence orthogonal to~${\mathcal H}_{\scriptscriptstyle{\text{\rm  com}}}$, so is the weak limit~$w_{\xi *}$. Proceeding as in (3), it can be shown that this $w_{\xi *}$  also is the strong limit of a sequence of elements of~${\mathcal H}_{r\rceil}$, hence belongs to ${\mathcal H}_{\scriptscriptstyle{\text{\rm   com}}}$; moreover, since the convergence is strong, Var($w_{\xi *}$)=1. The intersection of~${\mathcal H}_{\scriptscriptstyle{\text{\rm   com}}}$ and ${\mathcal H}_{\scriptscriptstyle{\text{\rm   idio}}}$ being~$\{0\}$, this implies $w_{\xi *}=0$. This, however, is incompatible with Var($w_{\xi *}$)=1. Therefore, the number of    diverging eigenvalues of ${\boldsymbol\Gamma}\n_{\chi}$ cannot be  strictly less than $r$. The Weyl inequalities and the fact that ${\boldsymbol\Gamma}\n_X = {\boldsymbol\Gamma}\n_\chi +{\boldsymbol\Gamma}\n_\xi$ then imply that all eigenvalues of ${\boldsymbol\Gamma}\n_\xi$ are bounded as~$n\to\infty$. Claims (ii) and (iii) follow.\hfill $\square$

\end{document}